\documentclass[journal = jpclcd, manuscript = letter]{achemso}

\usepackage{amsmath}
\usepackage{amssymb}
\usepackage{graphicx}
\usepackage{mathtools}
\usepackage{braket}
\usepackage{hyperref}
\usepackage{bbold}
\usepackage{calrsfs}
\usepackage{dsfont}
\usepackage{xcolor}
\usepackage{subcaption}

\newcommand{\iu}{{i\mkern1mu}}

\title{Chemistry in Quantum Cavities: Exact Results, the Impact of Thermal Velocities and Modified Dissociation}
  \author{Dominik Sidler}
  \email{dsidler@mpsd.mpg.de}
  \affiliation{Max Planck Institute for the Structure and Dynamics of Matter and Center for Free-Electron Laser Science \& Department of Physics, Luruper Chaussee 149, 22761 Hamburg, Germany}
  \author{Michael Ruggenthaler}
  \email{michael.ruggenthaler@mpsd.mpg.de}
  \affiliation{Max Planck Institute for the Structure and Dynamics of Matter and Center for Free-Electron Laser Science \& Department of Physics, Luruper Chaussee 149, 22761 Hamburg, Germany}
  \author{Heiko Appel}
  \email{heiko.appel@mpsd.mpg.de}
  \affiliation{Max Planck Institute for the Structure and Dynamics of Matter and Center for Free-Electron Laser Science \& Department of Physics, Luruper Chaussee 149, 22761 Hamburg, Germany}
  \author{Angel Rubio}
  \email{angel.rubio@mpsd.mpg.de}
  \affiliation{Max Planck Institute for the Structure and Dynamics of Matter and Center for Free-Electron Laser Science \& Department of Physics, Luruper Chaussee 149, 22761 Hamburg, Germany}
  \alsoaffiliation{Center for Computational Quantum Physics, Flatiron Institute, 162 5th Avenue, New York, NY 10010, USA}
  \alsoaffiliation{Nano-Bio Spectroscopy Group, Universidad del Pais Vasco, 20018 San Sebastian, Spain}

\begin{tocentry}
\centering
\includegraphics[width=1\linewidth]{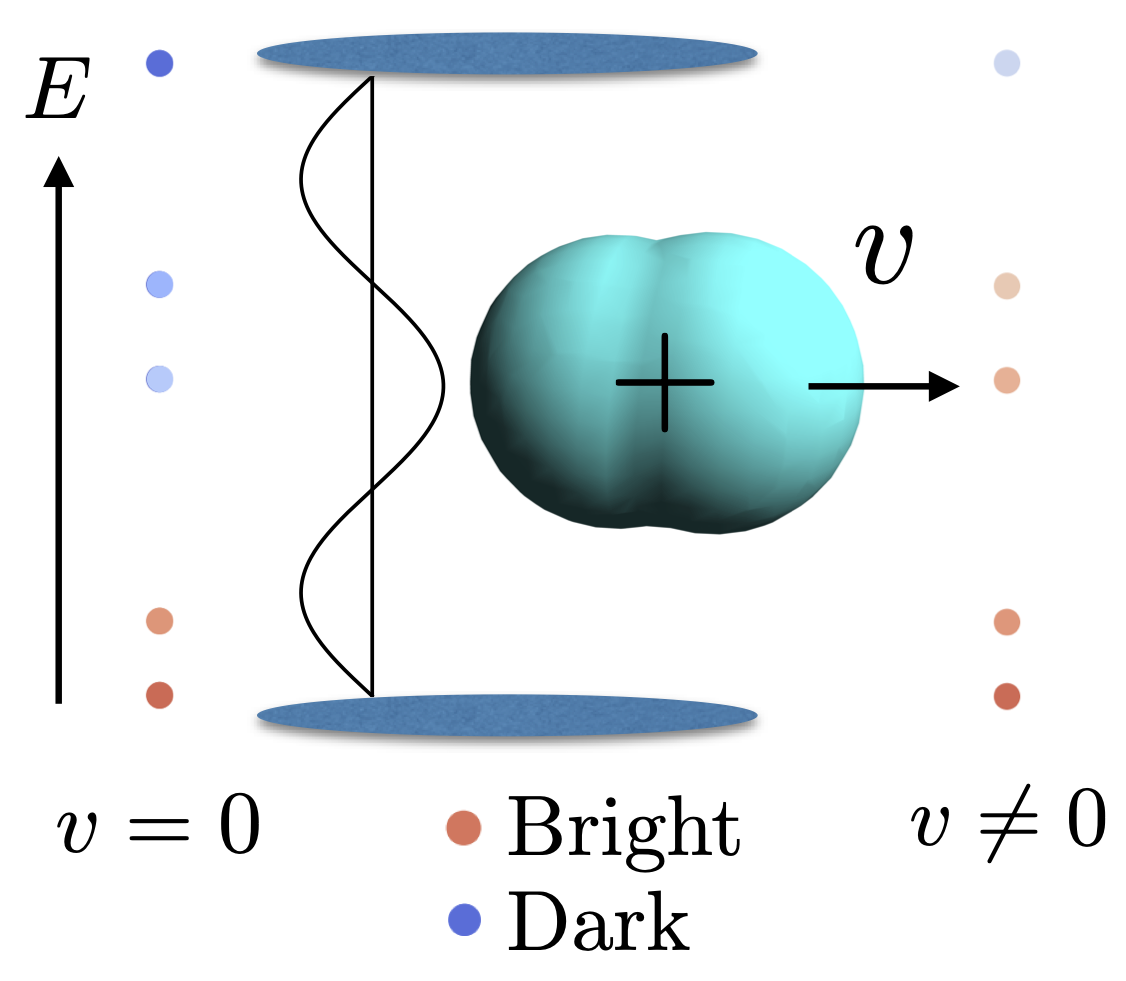}
\end{tocentry}


\begin{document}

\begin{abstract}
In recent years tremendous progress in the field of light-matter interactions has unveiled that strong coupling to the modes of an optical cavity can alter chemistry even at room temperature. Despite these impressive advances, many fundamental questions of chemistry in cavities remain unanswered. This is also due to a lack of exact results that can be used to validate and benchmark approximate approaches. In this work we provide such reference calculations from exact diagonalisation of the Pauli-Fierz Hamiltonian in the long-wavelength limit with an effective cavity mode. This allows us to investigate the reliability of the ubiquitous Jaynes-Cummings model not only for electronic but also for the case of ro-vibrational transitions. We demonstrate how the commonly ignored thermal velocity of charged molecular systems can influence chemical properties, while leaving the spectra invariant. Furthermore, we show the emergence of new bound polaritonic states beyond the dissociation energy limit.
\end{abstract}


\maketitle
 

Within the last few years, cavity-modified chemistry has gained popularity in the scientific community. This is due to several major breakthroughs in this emerging field of research~\cite{ebbesen2016,flick2017atoms,feist2017,ruggenthaler2018quantum,ribeiro2018polariton,flick2018strong}. For example, it was demonstrated that strong coupling in a cavity can be used to control reaction rates~\cite{hutchison2012,Thomas2016,schafer2019modification} or strongly increase energy-transfer efficiencies~\cite{coles2014polariton,zhong2017}. In contrast to the usual studies in quantum optics~\cite{kockum2019ultrastrong}, where ultra-high vacua and ultra-low temperatures are employed, many of these results were obtained at room temperature with relatively lossy cavities. Furthermore, recently it was even reported that strong coupling can modify the critical temperature of superconducting materials~\cite{sentef2018cavity,thomas2019exploring}. These results nurture the hope of the technological applicability of cavity-modified chemistry and material science.

Despite these experimental successes the understanding of the basic principles of cavity-modified chemistry (also called polaritonic or QED chemistry~\cite{flick2017atoms,schafer2019modification,flick2018ab,ruggenthaler2018quantum}) are still under debate~\cite{george2016multiple,herrera2016cavity,feist2015extraordinary,martinez2018can,doi:10.1021/acs.jpclett.0c00841,schafer2018ab,ruggenthaler2018quantum}. Currently, much of our understanding is based on quantum-optical models that have been designed for single (or a dilute gas of) atomic systems, whereas approximate first-principle simulations for coupled matter-photon situations emerge only slowly~\cite{flick2018ab,doi:10.1021/acsphotonics.9b00768,doi:10.1021/acs.jctc.7b00388,PhysRevLett.121.253001,triana2019revealing,csehi2019ultrafast,fregoni2020strong,doi:10.1080/00018732.2019.1695875}.
We believe it is pivotal to validate these model approaches and approximate first-principle simulations with numerically exact reference calculations to obtain a detailed understanding of cavity-modified chemistry and to see the limits of the different approximations used. Eventually, one should reach a level of certainty as it is the case in standard quantum chemistry~\cite{szabo2012}.

This work provides such references by presenting an exact-diagonalization scheme for the Pauli-Fierz Hamiltonian of non-relativistic quantum electrodynamics (QED) in the long-wavelength approximation for 3 interacting particles in 3D coupled to one effective photon mode. As examples, we present results for the He-atom and for HD+ and H$_2^+$ molecular systems in a cavity. We highlight that the inclusion of the quantized photons makes the interpretation of the obtained spectra much richer and more involved
, we discuss the level of accuracy of the ubiquitous Jaynes-Cummings model 
 and demonstrate fundamental effects beyond this model such as the formation of bound-states beyond the dissociation energy limit as well as the influence of the thermal velocity for charged systems. The latter point is specifically interesting, since it gives an indication why strong coupling has such an impact on chemistry at room temperature.

The consistent quantum description of photons coupled to matter is based on QED~\cite{ryder1996quantum,craig1998molecular,spohn2004}, which in its low-energy non-relativistic limit is given by the Pauli-Fierz Hamiltonian~\cite{spohn2004,ruggenthaler2014quantum}. For the case of optical and infrared wavelengths (dipole approximation) the Pauli-Fierz Hamiltonian in the Coulomb gauge can be further simplified~\cite{craig1998molecular,doi:10.1080/00018732.2019.1695875} and then reads in atomic units 
\begin{eqnarray}
\hat{H}&=&\sum_{i=1}^N \frac{\hat{\bold{p}}_{i}^2}{2m_i}+\sum_{i<j}^{N}\frac{Z_i Z_j}{|\hat{\bold{r}}_{i}-\hat{\bold{r}}_{j}|}+\sum_{\alpha=1}^{M_{pt}}\frac{1}{2}\bigg[\hat{p}_\alpha^2+\omega_\alpha^2\Big(\hat{q}_\alpha-\frac{\boldsymbol{\lambda}_\alpha}{\omega_\alpha}\cdot\hat{\bold{R}}\Big)^2\bigg].
\label{eq:PFLDip}
\end{eqnarray} 
Here $N$ is the number of charged particles (electrons and nuclei/ions) with mass $m_i$, charge $Z_i$ and $\hat{\bold{p}} = -\textrm{i} \boldsymbol{\nabla}$ is the non-relativistic momentum. Further, $M_{pt}$ photon modes with frequency $\omega_\alpha$ are coupled to the matter with the coupling $\boldsymbol{\lambda}_\alpha$ that contains the polarization vector and coupling strength of the individual modes. These couplings and frequencies are determined by the properties of the cavity. Further, $\hat{q}_\alpha$ and $\hat{p}_\alpha=-\textrm{i} \partial/\partial q_{\alpha}$ are the photon displacement and conjugate momentum operators, respectively, and the total dipole operator is defined as $\hat{\bold{R}}:=\sum_{i=1}^N Z_i \hat{\bold{r}}_i$.

\begin{figure}
    \centering
    \includegraphics[width=0.6\linewidth]{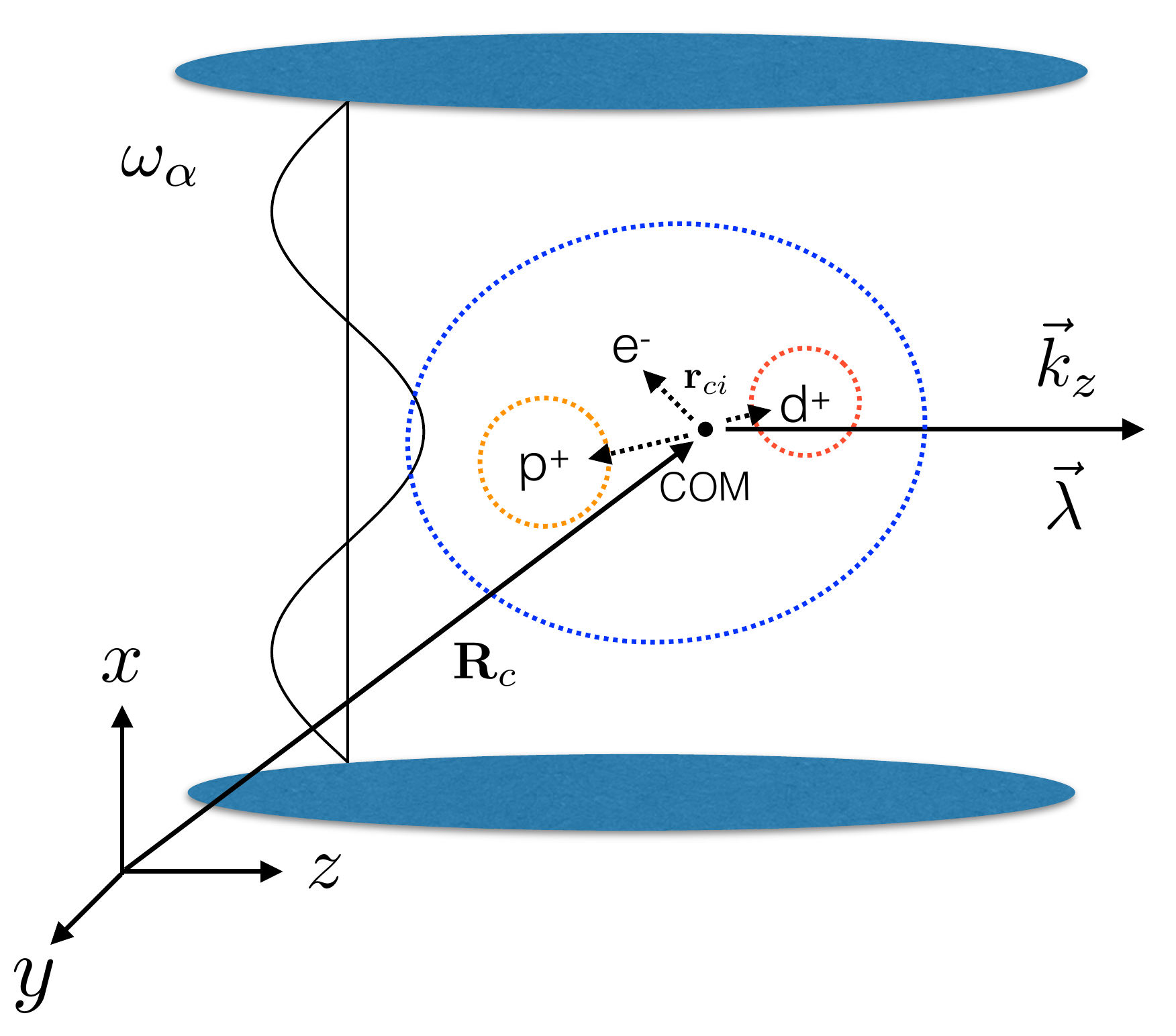}
    \caption{Schematics of the cavity-matter setup used here and exemplified for the HD+ molecule. We assume the relevant cavity mode polarized along the $z$-direction. The relative COM coordinates $\boldsymbol{r}_{ci}$ are given with respect to the COM $\boldsymbol{R}_{c}$.}
    \label{fig:hd_sketch}
\end{figure}

In the current work we will choose $N=3$ and consider the standard case of a single-mode cavity, i.e.\ $M_{pt}=1$, with polarization in $z$ direction (see Fig.~\ref{fig:hd_sketch}). To bring this numerically very challenging problem into a more tractable form we first re-express the Hamiltonian in terms of its centre-of-mass (COM) coordinates $\bold{r}_{ci}= \bold{r}_{i} - \bold{R}_c$, where the COM is given by $\bold{R}_c:= \tfrac{\sum_i m_i \bold{r}_i}{\sum_i m_i}$. Next we can shift the COM contribution of the total dipole operator to the COM momentum by a unitary Power-Zienau-Woolley transformation (see Sec.~1 of SI). The resulting eigenvalue equation can be brought into the form
\begin{eqnarray}
\Bigg[\frac{1}{2M}\bigg\{\bold{k}^2+\frac{2Q_{\textrm{tot}}\bold{k}\cdot \boldsymbol{\lambda}}{\omega^\prime}\hat p^\prime\bigg\} +\sum_{i=1}^3 \frac{\hat{\bold{p}}_{ci}^2}{2m_i}+\sum_{i<j}^{3}\frac{Z_i Z_j}{|\hat{\bold{r}}_{ci}-\hat{\bold{r}}_{cj}|}&&\nonumber\\
+\frac{1}{2}\bigg[\hat{p}^{\prime 2}+\omega^{\prime 2}\Big(\hat{q}^\prime-\frac{\boldsymbol{\lambda}}{\omega^\prime}\cdot\sum_{i=1}^3 Z_i \hat{\bold{r}}_{ci}\Big)^2\bigg]&\Bigg]  e^{\iu \bold{k}\boldsymbol{R}_c}\Phi^\prime
=&Ee^{\iu \bold{k}\boldsymbol{R}_c}\Phi^\prime,
\label{eq:PFHeig}
\end{eqnarray}
where we made the wave function ansatz 
$\psi^\prime = e^{\iu \bold{k}\boldsymbol{R}_c}\Phi^\prime$. Here $Q_\textrm{tot}:=\sum_i^3 Z_i$ is the total charge of the three particle system and we have performed a photon-coordinate transformation such that the frequency of the cavity becomes dressed $\omega^\prime=\omega\sqrt{1+\frac{1}{M}\Big( \frac{\lambda Q_{\textrm{tot}}}{\omega}\Big)^2}$. We already see that for charged systems, i.e.\ $Q_{\textrm{tot}} \neq 0$, we get novel contributions from the coupling of the COM motion with the quantized field that are not taken into account in usual quantum-optical models~\cite{kockum2019ultrastrong}. Therefore, in contrast to the usual Schr\"odinger equation, we will be able to show that the COM motion (corresponding to the continuous quantum number $\boldsymbol{k}$) has an influence on the bound states of the system. Such a contribution is to be expected, since moving charges will create a transversal electromagnetic field. Note that in the long-wavelength approximation this contribution only appears for charged systems, for the full (minimal-coupling) Pauli-Fierz Hamiltonian (i.e. beyond dipole approximation) small deviations are also expected for neutral systems~\cite{spohn2004}.

After separating off the COM coordinate with the above transformations, we can represent the three relative COM coordinates in terms of spherical-cylindrical coordinates~\cite{hesse2001lagrange}, i.e.\ $\bold{r}_{ci}(R,\theta,\phi,\rho,\psi,\zeta)$. Here $\zeta\in ]-\infty,\infty[$, $\{R,\rho\}\in [0,\infty[$ and the radial coordinates obey $\{\phi,\psi\}\in[0,2\pi[$ and $\theta\in[0,\pi[$. This allows us to express the wave function by
 \begin{eqnarray}
\Phi^\prime(\phi,\theta,\psi,R,\rho,\zeta,n)=\sum_{n=0}^{N_{pt}-1}\sum_{l,m=0}^{N_l,N_m}\sum_{k=-l}^lC_{l,m,n,k} D_{m,k}^l(\phi,\theta,\psi)\varphi_k(R,\rho,\zeta)\otimes\ket{n},
\label{eq:waveansatz}
\end{eqnarray} 
where $D_{m,k}^l$ are the Wigner-D-matrices~\cite{hesse2001lagrange}, with variational coefficients $C_{l,m,n,k}$. 
Photons are represented in a Fock number basis $\ket{n}$. Being formally exact, finite numerical precision is already indicated by the number of basis states $N_l,\ N_m,\ N_{pt}$ for the angular and photonic basis states. The radial wave function $\varphi_k$ is represented numerically in perimetric coordinates on a 3D Laguerre mesh~\cite{hesse2001lagrange} of dimensionality $N_{\textrm{matter}}^3$. Corresponding numerical parameters are given in Sec.~3 of the SI. In practice, radial integrals are solved numerically by a Gaussian quadrature, whereas angular integrals are solved analytically. We note that for an uncoupled setup (i.e. $\boldsymbol{\lambda}=0$) $m$ and $l$ correspond to the usual magnetic and angular quantum number, respectively.   
In this case, the expansion of Eq.~\eqref{eq:waveansatz} becomes highly efficient since the Hamiltonian assumes a block diagonal shape and it can be solved for each pair $m$ and $l$ independently (reducing the dimension of the problem to 3)~\cite{hesse1999lagrange,hesse2001lagrange,Rene_diploma}. Further simplifications can be made based on the parity invariance of the uncoupled problem. 
For our coupled problem, however, these symmetries are broken. Yet, due to the choice of the polarization direction we preserve cylindrical symmetry with respect to the lab frame and it can be shown that $\bra{\Phi^\prime_{l^\prime,m^\prime,n^\prime}} H^\prime \ket{\Phi^\prime_{l,m,n}} = \delta_{m^\prime,m}\bra{\Phi^\prime_{l^\prime,m^\prime,n^\prime}} H^\prime \ket{\Phi^\prime_{l,m,n}}$. 
Hence the coupling only mixes angular momenta and Fock states, which implies that the original 10-dimensional problem can be reduced to 6 dimensions. Different spin-states can be distinguished by (anti-)symmetrization of the matter-only wave functions. This is possible because we have at most 2 indistinguishable particles for bound 3-body problems and the Pauli-Fierz Hamiltonian in the long-wavelength limit is spin-independent. For further theoretical details see Sec.~2 as well as convergence tests see Sec.~3 of the SI. Note that our exact diagonalisation approach might also be suitable to investigate chiral cavities \textit{ab initio} with only minor modifications. They offer promising perspectives to control material properties by breaking time reversal symmetry.\cite{Huebner2020}

\begin{figure}
    \centering
    \includegraphics[width=0.7\linewidth]{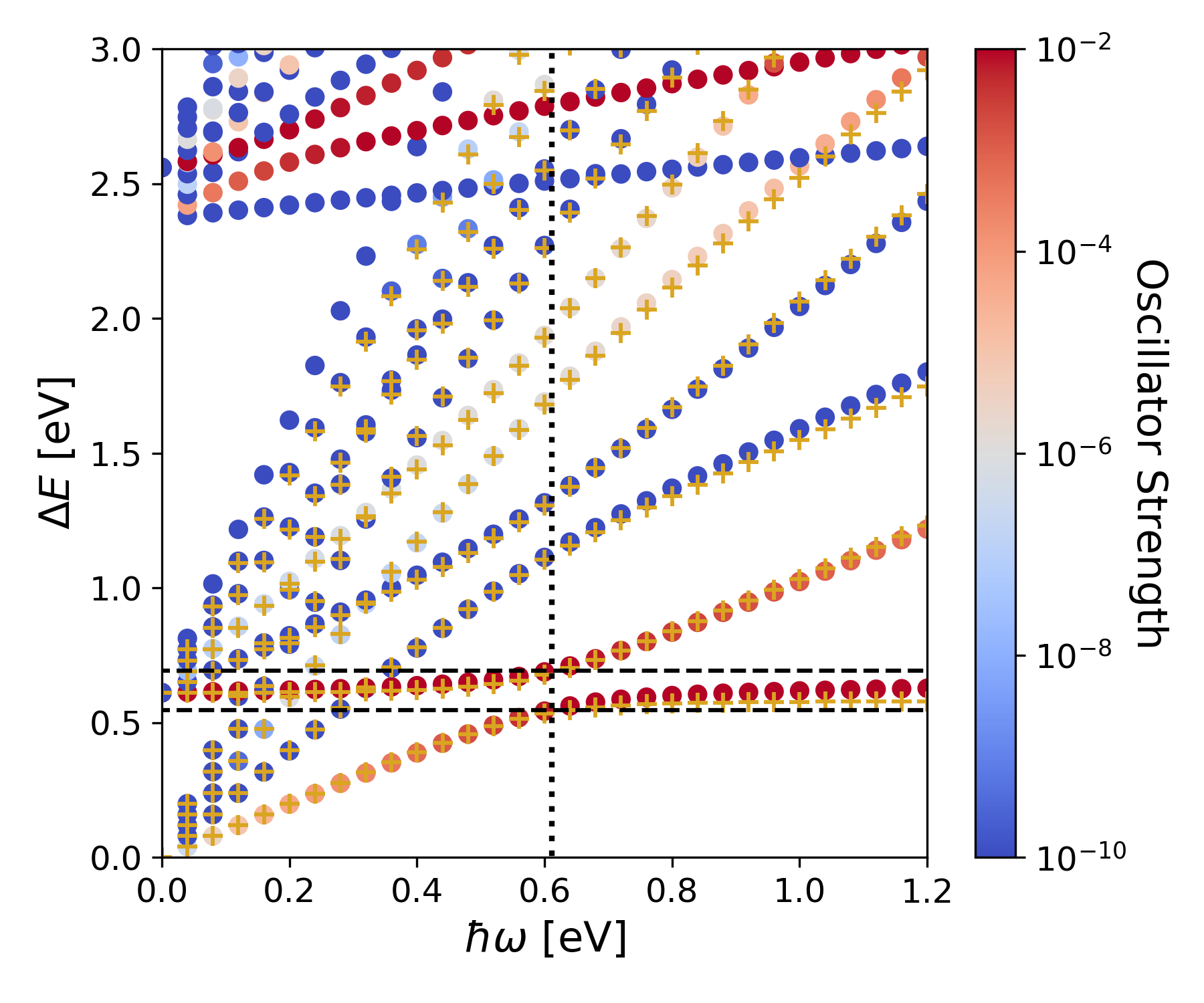}
    \caption{Rabi dispersion relation for parahelium in a cavity. The bright polaritonic states, located at the $\Delta E=E_{2P}-E_{2S}$ resonance frequency (vertical line) of the uncoupled system, are indicated by the two horizontal lines. They are associated with a high (red) dipole transition oscillatory strength, while corresponding dark many-photon replicas and improbable $2$S-$i$S transitions have a small oscillator strength (blue). The yellow crosses (+) indicate energies derived from the JC model based on an $2$S and $2$P two-level approximation and using the respective parameters for the cavity detuning frequencies and photon mode numbers.}
    \label{fig:disp_He}
\end{figure}

After having discussed how we exactly solve the problem of real systems coupled to the photons of an optical cavity numerically exactly, let us turn to the obtained results. As a first example we consider parahelium coupled to a cavity. We perform a scan of different frequencies $\omega$ centered around the $2$S-$2$P resonance coupling strength $g=0.074$ eV. Notice that when varying $\omega$, $\sqrt{\omega}/\lambda$ was kept constant for all calculations, i.e. $g\propto \omega$, and thus different coupling strengths were scanned trough at the same time. The first observation that can be made in the dispersion relation of Fig.~\ref{fig:disp_He} is that the spectrum of the Pauli-Fierz Hamiltonian becomes more intricate when compared to the usual Schr\"odinger Hamiltonian. The reason being that to each matter excitation we get photon replica spaced by roughly the corresponding photon frequency. This can be best observed for small frequencies, where we see clusters of eigenenergies. In our case we get 5 replica, where we have chosen the number of photon states $N_{pt}=6$. However, in principle we would get infinitely many discrete replicas at higher energies, which is an indication of the photon continuum. Moreover, if we simulated many modes, one would observe a continuum of energies starting at the ground state~\cite{spohn2004,doi:10.1021/acsphotonics.9b00768}. This photon continuum is necessary to capture fundamental physical processes like spontaneous emission and dissipation~\cite{doi:10.1021/acsphotonics.9b00768}, but it makes the identification of excited states difficult (in full QED they turn into resonances~\cite{ryder1996quantum, spohn2004}). That is why we have supplemented the energies in Fig.~\ref{fig:disp_He} with their color-coded oscillator strengths. This allows to associate the eigenenergies with large oscillator strengths to genuine resonances, i.e.\ they correspond to excited states with a finite line width. In a many-mode case the photon replica with smaller oscillator strength then constitute this linewidth~\cite{doi:10.1021/acsphotonics.9b00768}. At the $2$S-$2$P transition (indicated with a vertical line) we find a Rabi splitting $\Omega=35.78$ THz into the upper and lower polariton (indicated with two horizontal lines), which is of the order of $\Omega/\omega\sim 0.24$, hence we are in the strong-coupling regime~\cite{kockum2019ultrastrong}. Furthermore, we have indicated the predictions from the ubiquitous Jaynes-Cummings (JC) model\cite{jaynes1963comparison} based on the bare $2$S and $2$P states with yellow crosses. Since this model was constructed for atomic transitions on resonance it captures the Rabi splitting quite accurately, but for larger detuning parameters (i.e. off-resonance) it becomes less reliable. The JC model also gives a good approximation to the multi-photon replicas. However, since the JC model takes into account only the $2$S and $2$P bare-matter states in our case, all the other excitations are not captured.

\begin{figure}
\begin{subfigure}{.8\textwidth}
\centering
    \includegraphics[width=1\linewidth]{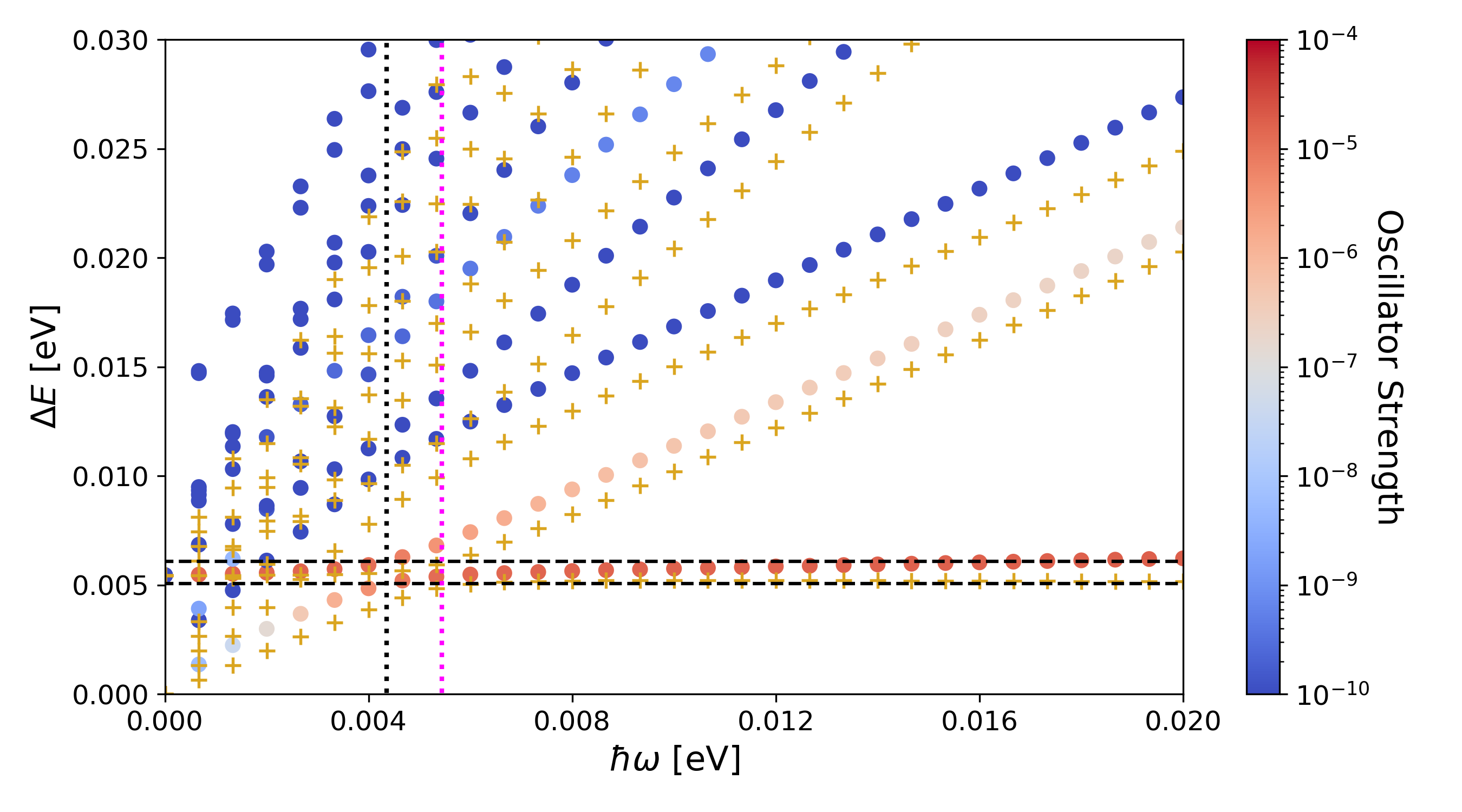}
    \caption{}
    \label{fig:disp_HDp_k0}
\end{subfigure}
\begin{subfigure}{.8\textwidth}
\centering
    \includegraphics[width=1\linewidth]{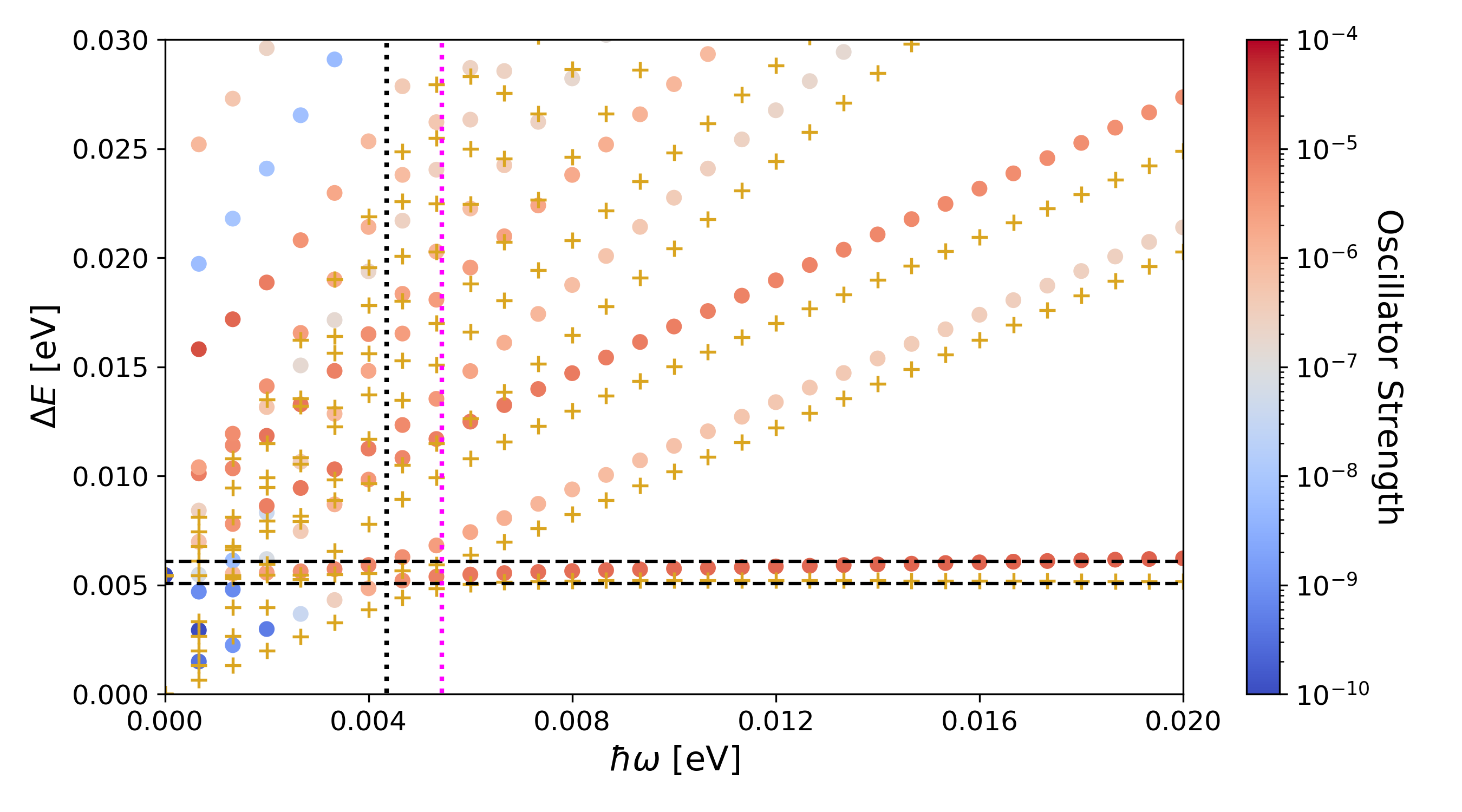}
    \caption{}
    \label{fig:disp_HDp_k1}
\end{subfigure}

\caption{In (a) and (b) we show the Rabi dispersion relation for HD+ in a cavity for COM motion $E_{kin}=0$ and $E_{kin}=0.24$e-2 eV, respectively. Similar results are discussed for the H$_2^+$ molecule in the SI. The bright polaritonic states at the dressed L$0$-L$1$ transition (vertical black line) are indicated by the two horizontal lines. The magenta vertical line shows the prediction of the JC model, which does not account for the net-charge frequency dressing. Dark (blue) and bright (red) states can be identified by corresponding dipole oscillator strengths. The yellow crosses (+) indicate energies derived from the JC model.  
}
\end{figure}

Let us switch from the atomic to the molecular case and consider transitions due to the nuclear motion. We here consider the HD+ molecule and the lowest ro-vibrational L$0$-L$1$ transition with a Rabi splitting of $\Omega=0.24$ THz. A similar dispersion plot as previously given for He can be seen in Fig.~\ref{fig:disp_HDp_k0}. The first difference is that we now have two vertical lines. The black vertical line corresponds to the (now dressed) resonance frequency $\omega$ of the system. The charged molecule slightly shifts the frequency of the empty cavity. The JC model, which does not take into account this effect, predicts the resonance at the magenta vertical line. In the HD+ case, where we find the exact value $\Omega/\omega \sim 0.23$, the JC model predicts instead a value of $0.28$ with the a wrong Rabi splitting of $0.36$ THz. In addition, the JC model underestimates polaritonic energy levels for all evaluated cavity frequencies in the ro-vibrational regime. This relatively strong deviation is due to the missing dipole self-energy term in the JC model and it highlights that few level atomic quantum-optical models are in principle less reliable when applied to molecular systems (see also Sec.~5 in the SI). As already anticipated in the theory part, the COM momentum in $z$ direction will have an influence on the eigenstates of HD+, since we consider a charged system. Indeed, while Fig.~\ref{fig:disp_HDp_k0} was calculated for zero momentum, in Fig.~\ref{fig:disp_HDp_k1} we see the dispersion plot for a finite COM kinetic energy $\frac{k_{z}^2}{2 M}=0.24$e-2 eV $\propto T=28.66$ K. Interestingly, the spectrum itself does not change, yet the eigenfunctions do so (additional information is provided in Sec.~5 in the SI). Consequently, previously dark transitions (small oscillator strength, blue) become bright (large oscillator strength, red). Therefore, the absorption/emission spectra, which depend on the oscillator strength, gets modified due to this COM motion and excitations to higher-lying states become more probable. Overall, the effect of the finite COM momentum appears to be strong for the infrared energy range. Note that we find similar results for H$_2^+$, which is shown in Sec.~6 in the SI. Since for realistic situations we will always have a thermal velocity distribution, these spectral modifications will become important. Specifically when we think about chemical reactions, where the properties of charged subsystems are essential, these modifications could help to explain the so-far elusive understanding of cavity-modified chemistry at room temperature.

\begin{figure}
    \centering
    \includegraphics[width=0.7\linewidth]{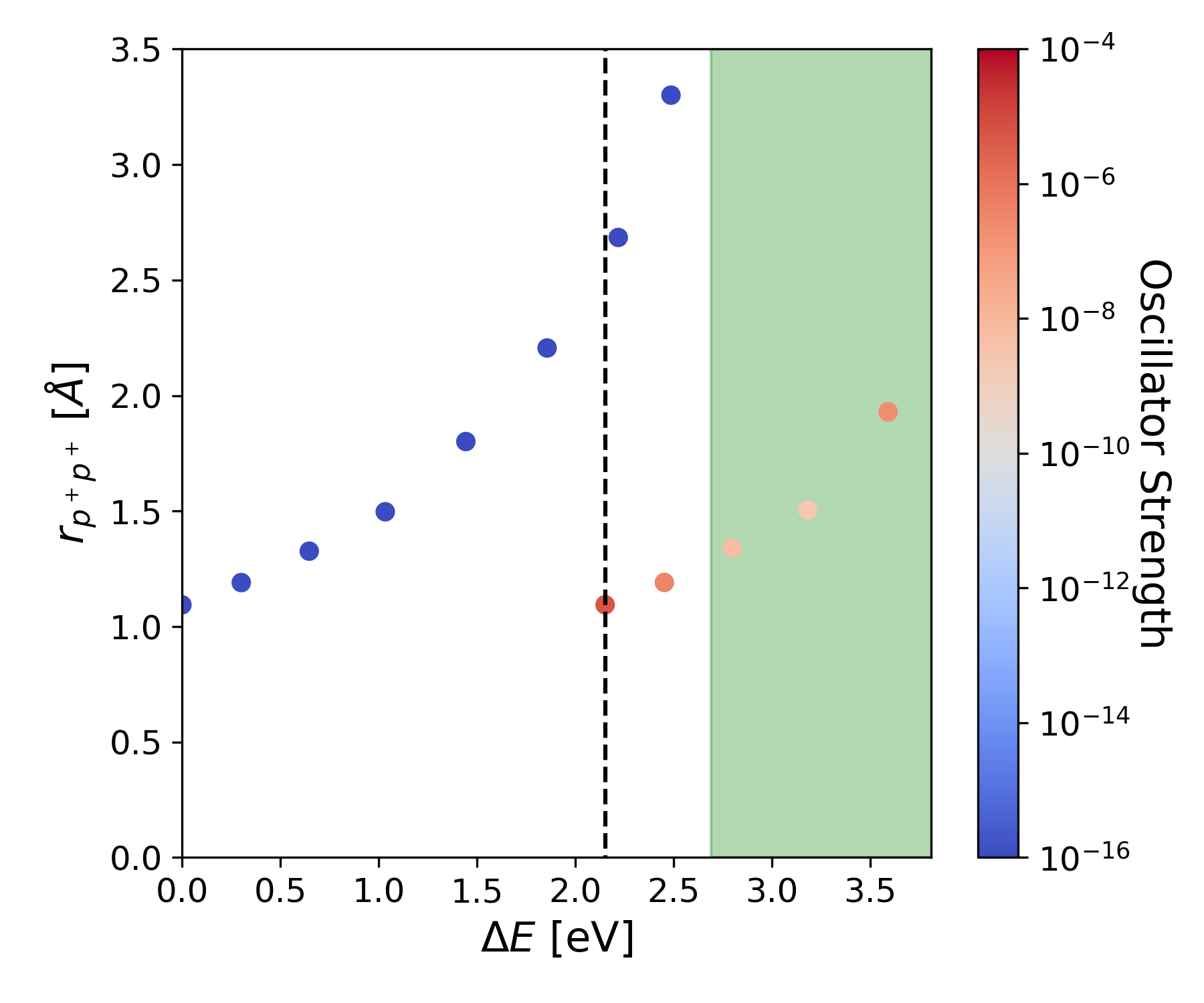}
    \caption{Quantized (i.e. bound) proton-proton distances for H$_2^+$ with respect to ground-state energy differences $\Delta E$ and corresponding oscillator strength. The blue dots correspond to dressed bare matter states whereas red dots indicate the emerging bright photon replicas, which are absent without a cavity. The cavity frequency is $\omega=2.15$ eV (dashed vertical line) with $\lambda=0.051$ and zero COM motion. The green area indicates energy ranges beyond the p$^+$-dissociation limit according to matter-only simulations. }
    \label{fig:dist_H2p}
\end{figure}

Another interesting result with relevance for polaritonic chemistry is the formation of bound polaritonic states below\cite{cortese2019strong,cortese2019excitons} and above the proton dissociation limit of H$_2^+$ (see green region in Fig.~\ref{fig:dist_H2p}). 
Since we treat the nuclei/ions quantum-mechanically, we do not have to approximate the Born--Oppenheimer surfaces in our present approach for a simple picture of dissociation. Therefore, we can identify the dissociation energy limit and the emergence of novel bound polaritonic states based on the expectation value of the proton-proton distance and by variation of the finite numerical grid (see Sec.~5 in SI). It is important to note that there are no dipole-allowed transitions to excited bound states available for the uncoupled case of  H$_2^+$, i.e. there are only $S$-type many-body eigenstates below the proton dissociation energy limit. Hence, if we couple  to the cavity with a frequency close to the dissociation energy, e.g. $\omega=2.15$ eV and $\lambda=0.051$, the Rabi model breaks down and no Rabi splitting is observed. Yet, while we find (dark) 
$S$-type states (blue dots) that follow the expected matter-only dissociation, multiple bright bound states (red dots) emerge, which can persist beyond the proton dissociation energy limit.
These states, which are bright photon replica of the bound matter-only $S$-states, employ the captured photons to bind the otherwise dissociating molecule.  How strong these states influence the molecular dissociation process has to be investigated in more detail in the future. It will depend also on whether they correspond to long-lived excited states or rather short-lived meta-stable states.

In this work we have provided numerically exact references for cavity-modified chemistry and we have demonstrated that the thermal velocity has a direct impact on properties of charged systems, as well as the emergence of bound polaritonic states beyond the dissociation-energy limit. We have done so by an exact diagonalization of the Pauli-Fierz Hamiltonian for 3 particles and one mode in center-of-mass coordinates and used further symmetries to reduce the originally 10-dimensional problem to a 6-dimensional one. We have shown that the resulting spectrum shows the onset of the photon continuum and hence is no longer obvious to interpret. Furthermore, for ro-vibrational transitions we have shown that the ubiquitous Jaynes-Cummings model is not very accurate and that for charged systems important properties like the oscillator strength are modified for non-zero center-of-mass motion. Since this can be connected to the thermal velocity, we found a so-far neglected contribution for cavity-modified chemistry at finite temperature. All these results highlight that at the interface between quantum optics and quantum chemistry well-established "common knowledge" is no longer necessarily applicable. In order to get a basic understanding of polaritonic chemistry and material science we need to revisit standard results and establish possibly new scientific facts, and numerically exact calculations of the basic QED equations are an integral part of this endeavor.

\begin{acknowledgement}
The authors thank Davis Welakuh, Christian Sch\"afer and Johannes Flick for helpful discussions and critical comments. In addition, many thanks to Rene Jest\"adt for providing his matter-only code, which acts as an invaluable basis for the implementation of the coupled problem. This work was made possible through the support of the RouTe Project (13N14839), financed by the Federal Ministry of Education and Research (Bundesministerium für Bildung und Forschung (BMBF)) and supported by the European Research Council (ERC-2015-AdG694097), the Cluster of Excellence "Advanced Imaging of Matter"(AIM) and Grupos Consolidados (IT1249-19). The Flatiron Institute is a division of the Simons Foundation.

\end{acknowledgement}

\begin{suppinfo}

\end{suppinfo}
\bibliography{ms}
\end{document}



\maketitle
 

\section{Theoretical Details}

\subsection{Center of Mass (COM) Separation}

In order to tackle the coupled light-matter problem defined by the Pauli-Fierz Hamiltoninan given in Eq.~(1) in dipole approximation, it is useful to switch to a centre-of-mass (COM) coordinate system. This means that we define $\bold{r}_i=\bold{R}_c+\bold{r}_{ci}$, with the COM explicitly given as $\bold{R}_c:= \frac{\sum_i m_i \bold{r}_i}{\sum_i m_i}$. The Hamiltonian can then be re-written in the COM frame as,

\begin{eqnarray}
\hat{H}&=&\frac{\hat{\bold{P}}_{c}^2}{2M}+\sum_{i=1}^N \frac{\hat{\bold{p}}_{ci}^2}{2m_i}+\sum_{i<j}^{N}\frac{ Z_i Z_j}{|\hat{\bold{r}}_{ci}-\hat{\bold{r}}_{cj}|}+\sum_{\alpha=1}^M\frac{1}{2}\bigg[\hat{p}_\alpha^2+\omega_\alpha^2\Big(\hat{q}_\alpha-\frac{\boldsymbol{\lambda}_\alpha}{\omega_\alpha}\cdot\hat{\bold{R}}\Big)^2\bigg].
\label{eq:PFHcom}
\end{eqnarray} 
Note that \textit{a priori} the dipole operator still contains a COM dependence at this stage. In a next step, we apply the unitary Power-Zienau-Woolley (PZW) transformation shifting the COM,
\begin{eqnarray}
\hat{U}_{PZW}&:=&e^{\iu \frac{Q_{\mathrm{tot}}\boldsymbol{\lambda}_\alpha\cdot \hat{\bold{R}}_c}{\omega_\alpha} \hat{p}_\alpha},
\label{eq:PFHcom2}
\end{eqnarray} 
with $Q_{\mathrm{tot}}:=\sum_{i=1}^N e Z_i$. By using the Baker-Campbell-Hausdorf formula, one can show that our PZW transformation obeys the following properties,
\begin{eqnarray}
\hat{U}_{PZW}\hat p_\alpha \hat{U}_{PZW}^\dagger &=& \hat p_\alpha\\
\hat{U}_{PZW}\hat q_\alpha \hat{U}_{PZW}^\dagger &=& \hat q_\alpha+\iu\frac{ Q_{\mathrm{tot}}\boldsymbol{\lambda}_\alpha\cdot \hat{\bold{R}}_c[\hat p_\alpha,\hat q_\alpha] }{\omega_\alpha}  \\
\hat{U}_{PZW}\hat{\bold{R}}_c \hat{U}_{PZW}^\dagger &=& \hat{\bold{R}}_c\label{eq:pzw_Rc}\\
\hat{U}_{PZW}\hat{\bold{P}}_c \hat{U}_{PZW}^\dagger& =& \hat{\bold{P}}_c+\iu \frac{Q_{\mathrm{tot}}\boldsymbol{\lambda}_\alpha\cdot [\hat{\bold{R}}_c,\hat{\bold{P}}_c]}{\omega_\alpha} \hat{p}_\alpha,
\end{eqnarray} 
which allow to write the shifted Hamiltonian as,
\begin{eqnarray}
\hat{H}^\prime&=&\frac{1}{2M}\bigg(\hat{\bold{P}}_{c}-\sum_{\alpha=1}^M\frac{\lambda_\alpha Q_{\textrm{tot}}}{\omega_\alpha}\hat p_\alpha\bigg)^2+\sum_{i=1}^N \frac{\hat{\bold{p}}_{ci}^2}{2m_i}+\sum_{i<j}^{N}\frac{Z_i Z_j}{|\hat{\bold{r}}_{ci}-\hat{\bold{r}}_{cj}|}\\
&&+\sum_{\alpha=1}^M\frac{1}{2}\bigg[\hat{p}_\alpha^2+\omega_\alpha^2\Big(\hat{q}_\alpha-\frac{\boldsymbol{\lambda}_\alpha}{\omega_\alpha}\cdot\sum_{i=1}^N Z_i \hat{\bold{r}}_{ci}\Big)^2\bigg],
\label{eq:PFHshifted}
\end{eqnarray}
by applying the PZW transformation for each mode $\alpha$.
Afterwards, the original eigenvalue problem $H\psi=E\psi$ can be simplified to Eq.~(2) given in the letter by using a wave function Ansatz of the form $\psi^\prime(R_c,r_c,q_\alpha)=e^{\iu \bold{k}\boldsymbol{R}_c}\Phi^\prime(r_c,q_\alpha)$. For one mode $\alpha$ and three bodies, one obtains the following expression,

\begin{eqnarray}
\Bigg[\frac{1}{2M}\bigg\{\bold{k}^2+\frac{2Q_{\textrm{tot}}\bold{k}\cdot \boldsymbol{\lambda}}{\omega^\prime}\hat p^\prime\bigg\} +\sum_{i=1}^3 \frac{\hat{\bold{p}}_{ci}^2}{2m_i}+\sum_{i<j}^{3}\frac{Z_i Z_j}{|\hat{\bold{r}}_{ci}-\hat{\bold{r}}_{cj}|}&&\nonumber\\
+\frac{1}{2}\bigg[\hat{p}^{\prime 2}+\omega^{\prime 2}\Big(\hat{q}^\prime-\frac{\boldsymbol{\lambda}}{\omega^\prime}\cdot\sum_{i=1}^3 Z_i \hat{\bold{r}}_{ci}\Big)^2\bigg]&\Bigg]  e^{\iu \bold{k}\boldsymbol{R}_c}\Phi^\prime
=&Ee^{\iu \bold{k}\boldsymbol{R}_c}\Phi^\prime.
\label{eq:PFHeig}
\end{eqnarray}
In case of neutral systems (i.e. $Q_{\textrm{tot}}=0$), the additional interaction of the COM motion with the quantized field vanishes.
Note that in Eq.~(\ref{eq:PFHeig}) it was used that for one mode (i.e. $M=1$) the PZW-shifted eigenvalue problem can be additionally simplified by absorbing $\frac{1}{M}\Big( \frac{\lambda Q_{\textrm{tot}}}{\omega}\Big)^2$ in a dressed resonance frequency, 
$\omega^\prime=\omega\sqrt{1+\frac{1}{M}\Big( \frac{\lambda Q_{\textrm{tot}}}{\omega}\Big)^2}$ with the corresponding momenta $\hat{p}^\prime=\iu\sqrt{\frac{\omega^\prime}{2}}(\hat{a}^\dagger-\hat{a})$ and position operators $\hat{q}^\prime=\sqrt{\frac{1}{2\omega^\prime }}(\hat{a}^\dagger+\hat{a})$. They obey the usual canonical commutation relations $[\hat{q}^\prime,\hat{p}^\prime]=\iu$. 
This shift will only be relevant for relatively light charged particles at low resonance frequencies $\omega$ (e.g. fundamental L$0$-L$1$ transition of HD+). 

\subsubsection{Observables}

When calculating observables of the coupled system ($\lambda\neq 0$), \textit{a priori} one cannot neglect any involved coordinate. However, in practise not always all integrals have to be solved explicitly.
For example, the dipole oscillatory strengths can be calculated from COM relative coordinates $\textbf{r}_{ci}$ as follows:
\begin{eqnarray}
\mathrm{Osc}_{jk}&=&\frac{2}{\frac{1}{m_1}+\frac{1}{m_2}+\frac{1}{m_3}}(E_j-E_k)|\bra{\psi_j}\hat{\textbf{R}}\otimes \mathds{1}_{pt}\ket{\psi_k}\nonumber \\
&=&\frac{2}{\frac{1}{m_1}+\frac{1}{m_2}+\frac{1}{m_3}}(E_j-E_k)|\bra{\psi_j^\prime}\sum_{i=1}^3 Z_i e \hat{\textbf{r}}_{ci}\otimes \mathds{1}_{pt}\ket{\psi^\prime_k},
\end{eqnarray}
where in the last step it was used that $\bra{\psi}\hat{\textbf{R}}\otimes \mathds{1}_{pt}\ket{\psi}=\bra{\psi^\prime}\hat{U}_{PZW}(\hat{\textbf{R}}\otimes \mathds{1}_{pt})\hat{U}^\dagger_{PZW}\ket{\psi^\prime}=\bra{\psi^\prime}\hat{\textbf{R}}\otimes \mathds{1}_{pt}\ket{\psi^\prime}$ with Eq. (\ref{eq:pzw_Rc}).

In contrast, photonic observables (e.g. Mandel $Q$-parameter) have to be transformed back to the length gauge to be consistent. Hence, an integration over the COM position has to be performed explicitly. However, due to the cylindrically symmetric setup with respect to the $z$-axis of the lab frame, i.e.
\begin{eqnarray}
\bra{\psi}\mathds{1}_{\mathrm{matter}}\otimes\hat{O}_{pt}\ket{\psi}=\bra{\psi^\prime}\mathds{1}_{\mathrm{matter}}\otimes \hat{U}(Z_c)\hat{O}_{pt}\hat{U}^\dagger_{PZW}(Z_c)\ket{\psi^\prime},
\end{eqnarray}
the COM integration is reduced to one dimension only.

The general expectation-value integral in our chosen spherical-cylindrical coordinate system (see next section) is given as,
\begin{eqnarray}
\bra{\psi}\hat{O}\ket{\psi}&=&\bra{\psi^\prime}\hat{U}_{PZW}\hat{O}(p_{ci},r_{ci},\hat{p},\hat{q})\hat{U}_{PZW}^\dagger\ket{\psi^\prime}\nonumber\\
&=&\sum_{n=0}^\infty\int_{-\infty}^\infty dR_{cx}\int_{-\infty}^\infty dR_{cy}\int_{-\infty}^\infty dR_{cz}\int_{-\infty}^\infty d\zeta\int_{0}^\infty dR\int_{0}^\infty d\rho\int_0^{2\pi} d\phi\int_0^{\pi} d\theta\int_0^{\pi}d \psi \nonumber\\
&&R^2\rho \sin\theta e^{-\iu k_z R_{cz}}\Phi^{\prime *} e^{\iu \frac{Q_\textrm{tot}\lambda R_{cz}\hat p}{\omega}}\hat{O}e^{-\iu \frac{Q_\textrm{tot}\lambda R_{cz}\hat p}{\omega}} e^{\iu k_z R_{cz}}\Phi^\prime\\
&=&\sum_{n=0}^\infty \int_{-\infty}^\infty dR_{cz}\int_{-\infty}^\infty d\zeta\int_{0}^\infty dR\int_{0}^\infty d\rho\int_0^{2\pi} d\phi\int_0^{\pi} d\theta\int_0^{\pi}d \psi \\
&&R^2\rho \sin\theta \Phi^{\prime *}e^{\iu \frac{Q_\textrm{tot}\lambda R_{cz}\hat p}{\omega}}\hat{O}e^{-\iu \frac{Q_\textrm{tot}\lambda R_{cz}\hat p}{\omega}} \Phi^\prime,
\end{eqnarray}
where in the last step it was assumed that $\hat O$ does not explicitly depend on the COM coordinate and positions. Note, if $\lambda=0$ or if $Q_\textrm{tot}=0$ or if $\hat O$ does not depend on $\hat{q}$ and $P_{cz}$, the $R_{cz}$ integral is unity as it was the case for a matter only system.

\subsection{Basis Set Representation}
First of all, $\lambda$-coupling is assumed along the z-axis only, i.e. 
\begin{align}
  \boldsymbol{\lambda}_\alpha=\begin{bmatrix}0\\0\\ \lambda_\alpha\end{bmatrix}  .
\end{align}
For 3 bodies, it is helpful to express the relative coordinates of the  centre of mass frame in a combined spherical-cylindrical coordinate system, i.e. $\bold{r}_{ci}(R,\theta,\phi,\rho,\psi,\zeta)$ with ${\zeta}\in [-\infty,\infty[$,$\{R,\rho\}\in [0,\infty[$, $\{\phi,\psi\}\in[0,2\pi[$ and $\theta\in[0,\pi[$. The particle vectors are explicitly given as~\cite{hesse2001lagrange,Rene_diploma}

\begin{eqnarray}
\bold{r}_{c1}=\begin{bmatrix}x_{c1}\\y_{c1}\\z_{c1}\end{bmatrix}=\begin{bmatrix}\frac{m_2+m_3/2}{M}R\sin{\theta}\cos{\phi}+\frac{m_3}{M}\big((\rho \cos{\theta}\cos{\psi}+\zeta\sin{\theta})\cos{\phi}-\rho\sin{\psi}\sin{\phi}\big)\\
\frac{m_2+m_3/2}{M}R\sin{\theta}\sin{\phi}+\frac{m_3}{M}\big((\rho \cos{\theta}\cos{\psi}+\zeta\sin{\theta})\sin{\phi}-\rho\sin{\psi}\cos{\phi}\big)\\
\frac{m_2+m_3/2}{M}R\cos{\theta}+\frac{m_3}{M}\big(-\rho \sin{\theta}\cos{\psi}+\zeta\cos\big)\end{bmatrix}\label{eq:rc1}\\
\bold{r}_{c2}=\begin{bmatrix}x_{c2}\\y_{c2}\\z_{c2}\end{bmatrix}=\begin{bmatrix}-\frac{m_1+m_3/2}{M}R\sin{\theta}\cos{\phi}+\frac{m_3}{M}\big((\rho \cos{\theta}\cos{\psi}+\zeta\sin{\theta})\cos{\phi}-\rho\sin{\psi}\sin{\phi}\big)\\
-\frac{m_1+m_3/2}{M}R\sin{\theta}\sin{\phi}+\frac{m_3}{M}\big((\rho \cos{\theta}\cos{\psi}+\zeta\sin{\theta})\sin{\phi}-\rho\sin{\psi}\cos{\phi}\big)\\
-\frac{m_1+m_3/2}{M}R\cos{\theta}+\frac{m_3}{M}\big(-\rho \sin{\theta}\cos{\psi}+\zeta\cos\big)\end{bmatrix}\label{eq:rc2}\\
\bold{r}_{c3}=\begin{bmatrix}x_{c3}\\y_{c3}\\z_{c3}\end{bmatrix}=\begin{bmatrix}\frac{m_2-m_1}{2M}R\sin{\theta}\cos{\phi}-\frac{m_1+m_2}{M}\big((\rho \cos{\theta}\cos{\psi}+\zeta\sin{\theta})\cos{\phi}-\rho\sin{\psi}\sin{\phi}\big)\\
\frac{m_2-m_1}{2M}R\sin{\theta}\sin{\phi}-\frac{m_1+m2}{M}\big((\rho \cos{\theta}\cos{\psi}+\zeta\sin{\theta})\sin{\phi}-\rho\sin{\psi}\cos{\phi}\big)\\
\frac{m_2-m_1}{2M}R\cos{\theta}-\frac{m_1+m_2}{M}\big(-\rho \sin{\theta}\cos{\psi}+\zeta\cos\big)\end{bmatrix}.\label{eq:rc3}
\end{eqnarray} 
and the corresponding transformed volume element becomes
\begin{eqnarray}
dV=R^2\rho \sin(\theta) dR_{cx}dR_{cy}dR_{cz}d\zeta dR d\rho d\phi d\theta d\psi.
\label{eq:vol1}
\end{eqnarray} 
In a next step, we employ the Ansatz wave function defined in Eq.~(3) of the main text, 
 which gives access to the formally exact solution for $N_l,N_m,N_{pt}\rightarrow \infty$.
For uncoupled setups (i.e. $\lambda=0$), $l$ refers to the to the angular quantum number and $m$ to the magnetic quantum number, which describe the total angular momentum relation $L^2=l(l+1)$ and its $z$-projection $L_z=m$. 
 Suppose we want to restrict the magnetic quantum number $m$ to zero, which is a priori a reasonable choice for a matter only or uncoupled systems by setting $N_m=0$. 

 Due to the choice of $\boldsymbol{\lambda}\parallel z$, i.e. by preserving the cylindrical symmetry with respect to the $z$-axis of the lab frame, and by using the definition of Wigner D-Matrices $D_{m,k}^{j}=e^{-\iu m\phi}d^j_{m,k}(\theta) e^{-\iu k \psi}$ with Wigner's (small) d-matrix defined according to standard literature, one can show that $\bra{\Phi_{l^\prime,m^\prime}}  H^\prime_{pt}\ket{\Phi_{l,m}} = \delta_{m^\prime,m}\bra{\Phi_{l^\prime,m^\prime}}  H^\prime_{pt}\ket{\Phi_{l,m}}$, since $H^\prime_{pt}$ does not depend on $\phi$. In other words, restricting $m=0$ is a valid choice even for coupled systems. However, the coupling of the photons to the matter starts to mix angular states. Hence, one cannot diagonalize the coupled Hamiltonian anymore for each angular momentum quantum number $l$ separately, which increases the dimensionality of the coupled problem considerably apart from the extra photonic degree of freedom. 
For practical reasons (implementation amount and computational load), we restrict the basis size to S and P states only, i.e. $l<2$, for all subsequent calculations. Therefore, the Wigner-D matrix wave function Ansatz, given in Eq. (3) of the main text, can be rewritten in terms of superpositions of even (e) and odd (o) wave-functions of the P-states leading to the following orthonormal basis~\cite{hesse2001lagrange}
 \begin{eqnarray}
\Phi^\prime_S&:=&\frac{1}{\sqrt{8}\pi} \varphi_S(R,\rho,\zeta)\otimes\ket{n}\label{eq:waveansatz_first}\\
{\Phi^\prime_P}^{\mathrm{e}}&:=&\frac{\sqrt{3}}{\sqrt{8}\pi} \sin(\theta)\cos(\psi)\varphi_P^{\mathrm{e}}(R,\rho,\zeta)\otimes\ket{n}\\
{\Phi^\prime_{P0}}^{\mathrm{o}}&:=&\frac{\sqrt{3}}{\sqrt{8}\pi}\cos(\theta) \varphi_{P0}^{\mathrm{o}}(R,\rho,\zeta)\otimes\ket{n}\\
{\Phi^\prime_{P1}}^{\mathrm{o}}&:=&-\frac{\sqrt{3}}{\sqrt{8}\pi}\sin(\theta)\cos(\psi) \varphi_{P1}^{\mathrm{o}}(R,\rho,\zeta)\otimes\ket{n}.
\label{eq:waveansatz_last}
\end{eqnarray} 
The resulting representation of the Pauli-Fierz Hamiltonian takes the following block-diagonal form:
\begin{eqnarray}
\mathcal{H}^\prime&=&\mathcal{H}^\prime_{m}+\mathcal{H}^\prime_{pt}=\begin{bmatrix}H_{SS}(\hat{\textbf{p}}_{ci},\hat{\textbf{r}}_{ci})&0&0&0\\
0&H_{PP}(\hat{\textbf{p}}_{ci},\hat{\textbf{r}}_{ci})&0&0\\
0&0&H_{P0P0}(\hat{\textbf{p}}_{ci},\hat{\textbf{r}}_{ci})&H_{P0P1}(\hat{\textbf{p}}_{ci},\hat{\textbf{r}}_{ci})\\
0&0&H_{P1P0}(\hat{\textbf{p}}_{ci},\hat{\textbf{r}}_{ci})&H_{P1P1}(\hat{\textbf{p}}_{ci},\hat{\textbf{r}}_{ci})
\end{bmatrix}\nonumber\\
&&+\begin{bmatrix}H_{SS}(\hat{\textbf{r}}_{ci},\hat{p}^\prime,\hat{q}^\prime)&0&H_{SP0}(\hat{\textbf{r}}_{ci},\hat{q}^\prime)&H_{SP1}(\hat{\textbf{r}}_{ci},\hat{q}^\prime)\\
0&H_{PP}(\hat{\textbf{r}}_{ci},\hat{p}^\prime,\hat{q}^\prime)&0&0\\
H_{P0S}(\hat{\textbf{r}}_{ci},\hat{q}^\prime)&0&H_{P0P0}(\hat{\textbf{r}}_{ci},\hat{p}^\prime,\hat{q}^\prime)&H_{P0P1}(\hat{\textbf{r}}_{ci},\hat{q}^\prime)\\
H_{P1S}(\hat{\textbf{r}}_{ci},\hat{q}^\prime)&0&H_{P1P0}(\hat{\textbf{r}}_{ci},\hat{q}^\prime)&H_{P1P1}(\hat{\textbf{r}}_{ci},\hat{p}^\prime,\hat{q}^\prime)
\end{bmatrix}
\label{eq:bd_m}
\end{eqnarray} 
where the first term corresponds to the matter-only problem  promoted to the coupled matter-photon space, e.g. $H_{ij}=H_{ij}^m\otimes \mathds{1}_{\mathrm{pt}}$ with matrix elements $H_{ij}^m$ given in the literature~\cite{Rene_diploma}.
Note that vanishing matrix entries in the first term are due to parity symmetry of the uncoupled problem. Vanishing matrix entries in the second term are obtained by analytical angular integration in combination with the chosen basis set truncation at $l=1$. The matrix elements are explicitly given as

\begin{eqnarray}
H_{SP0,ij}=H_{P0S,ji}&=&-\frac{\sqrt{3}}{3}\omega\lambda\Bigg\langle \bigg\{ Z_1  \Big[\frac{m_1 + m_3 / 2}{M} R +
                                   \frac{m_3}{M} \zeta \Big]\\
                                   &&+ Z_2 \Big[-\frac{m_1 + m_3 / 2}{M} R +
                                   \frac{m_3}{M} \zeta \Big]
                                   + Z_3 \Big[\frac{m_2- m_1}{2M} R -\frac{m_1+m_2}{M}
                                   \zeta\Big]\bigg\} \hat{q}^\prime\Bigg\rangle_{ij}\nonumber\label{eq:mel_first}\\
H_{SP1,ij}=H_{P1S,ji}&=&\frac{\sqrt{3}}{3}\omega\lambda\Bigg\langle \bigg\{-Z_1\frac{m_3}{M} 
                                   -Z_2\frac{m_3}{M}
                                   + Z_3 \frac{m_1+m_2}{M} \bigg\}\rho \hat{q}^\prime\Bigg\rangle_{ij}\\
H_{SS,ij}&=&\Bigg\langle\frac{1}{2}\bigg\{\frac{k_z^2}{M}+\hat{p}^{\prime 2}+\omega^{\prime 2}\hat{q}^{\prime 2} +\frac{2 Q_{\mathrm{tot}} k_z \lambda}{M\omega^\prime}\hat{p}^\prime\bigg\}\Bigg\rangle_{ij}+\\
&& \frac{\lambda^2}{2}\Big\langle Z_1^2 z_{1c}^2+Z_2 z_{2c}^2+Z_3 z_{3c}^2+2(Z_1 Z_2 z_{1c}z_{2c}+Z_1 Z_3 z_{1c}z_{3c}+Z_2 Z_3 z_{2c}z_{3c})\Big\rangle_{ij}\nonumber\\
H_{PP,ij}&=&\Bigg\langle\frac{1}{2}\bigg\{\frac{k_z^2}{M}+\hat{p}^{\prime 2}+\omega^{\prime 2}\hat{q}^{\prime 2} +\frac{2 Q_{\mathrm{tot}} k_z \lambda}{M\omega^\prime}\hat{p}^\prime\bigg\}\Bigg\rangle_{ij}+\\
&& \frac{\lambda^2}{2}\Big\langle Z_1^2 z_{1c}^2+Z_2 z_{2c}^2+Z_3 z_{3c}^2+2(Z_1 Z_2 z_{1c}z_{2c}+Z_1 Z_3 z_{1c}z_{3c}+Z_2 Z_3 z_{2c}z_{3c})\Big\rangle_{ij}\nonumber\\
H_{P0P0,ij}&=&\Bigg\langle\frac{1}{2}\bigg\{\frac{k_z^2}{M}+\hat{p}^{\prime 2}+\omega^{\prime 2}\hat{q}^{\prime 2} +\frac{2 Q_{\mathrm{tot}} k_z \lambda}{M\omega^\prime}\hat{p}^\prime\bigg\}\Bigg\rangle_{ij}+\\
&& \frac{\lambda^2}{2}\Big\langle Z_1^2 z_{1c}^2+Z_2 z_{2c}^2+Z_3 z_{3c}^2+2(Z_1 Z_2 z_{1c}z_{2c}+Z_1 Z_3 z_{1c}z_{3c}+Z_2 Z_3 z_{2c}z_{3c})\Big\rangle_{ij}\nonumber\\
H_{P1P1,ij}&=&\Bigg\langle\frac{1}{2}\bigg\{\frac{k_z^2}{M}+\hat{p}^{\prime 2}+\omega^{\prime 2}\hat{q}^{\prime 2} +\frac{2 Q_{\mathrm{tot}} k_z \lambda}{M\omega^\prime}\hat{p}^\prime\bigg\}\Bigg\rangle_{ij}+\\
&& \frac{\lambda^2}{2}\Big\langle Z_1^2 z_{1c}^2+Z_2 z_{2c}^2+Z_3 z_{3c}^2+2(Z_1 Z_2 z_{1c}z_{2c}+Z_1 Z_3 z_{1c}z_{3c}+Z_2 Z_3 z_{2c}z_{3c})\Big\rangle_{ij}\nonumber\\
H_{P1P0,ij}=H_{P0P1,ji}&=& \frac{\lambda^2}{2}\Big\langle Z_1^2 z_{1c}^2+Z_2 z_{2c}^2+Z_3 z_{3c}^2+\\
&&2(Z_1 Z_2 z_{1c}z_{2c}+Z_1 Z_3 z_{1c}z_{3c}+Z_2 Z_3 z_{2c}z_{3c})\Big\rangle_{ij}\nonumber \label{eq:mel_last}
\end{eqnarray} 
with 
\begin{eqnarray}
 z_{1c}^2 &=& \bigg(\frac{m_2 + m_3 / 2 }{M}\bigg)^2 \beta+
            \frac{m_3 (m_2 + m_3 / 2)}{M^2}  \gamma + \bigg(\frac{m_3 }{ M}\bigg)^2 \epsilon\\
        z_{2c}^2 &=& \bigg(\frac{m_1 + m_3 / 2}{ M}\bigg)^2 \beta -
            \frac{m_3 (m_1 + m_3 / 2)}{M^2} \gamma + \bigg(\frac{m_3 }{ M}\bigg)^2  \epsilon2\\
        z_{3c}^2 &=& \bigg(\frac{m_2 - m_1} {2 M}\bigg)^2 \beta- \frac{(m_2 - m_1) 
            (m_1 + m_2)}{ 2M^2} \gamma+ \bigg(\frac{m_1 + m_2}{M}\bigg)^2 \epsilon\\
        z_{1c}z_{2c} &=& -\frac{m_2 + m_3 / 2}{M} \frac{m_1 + m_3 / 2}{M} \beta +
            \bigg(\frac{m_2 + m_3 / 2}{M} - \frac{m_1 + m_3 / 2}{ M}\bigg) \frac{m_3}{M} \gamma \nonumber\\
            &&+\bigg(\frac{m_3}{M}\bigg)^2 \epsilon\\
        z_{1c}z_{3c} &=& \frac{m_2 + m_3 / 2}{ M} \frac{m_2 - m_1}{ 2 M} \beta + 
            \bigg(-\frac{m_2 + m_3 / 2}{M} \frac{m_1 + m_2}{ M} +
             \frac{m_3(m_2 - m_1)}{2 M^2}\bigg)\gamma\nonumber\\
             && - \frac{m_3(m_1+m_2)}{M^2} \epsilon\\
        z_{2c}z_{3c} &=& -\frac{m_1 + m_3 / 2}{ M} \frac{m_2 - m_1}{2 M} \beta +
            \bigg(\frac{m_1 + m_3 / 2}{M} \frac{m_1 + m_2}{M} +
              \frac{m_3(m_2 - m_1)}{2M^2}\bigg) \gamma\nonumber\\
              &&- \frac{m_3(m_1+m_2)}{M^2} \epsilon,
\end{eqnarray}
where $\beta,\gamma,\epsilon$ are defined in spherical-cylindrical coordinates as,
\begin{eqnarray}
\beta&=& b_1 R^2 \\ 
\gamma&=&c_1 R\zeta-c_2R\rho\\ 
\epsilon&=&e_1 \rho^2+e_2 \zeta^2-2 e_3 \rho\zeta. 
\end{eqnarray}
The coefficients contain the analytical evaluation of the angular integrals of the $\lambda^2$-term, which amount to the following non-zero values,
\begin{eqnarray}
b_{1SS}&=&\frac{1}{3},\ b_{1PP}=b_{1P1P1}=\frac{3}{15},\ b_{1P0P0}=\frac{3}{5}\\
c_{1SS}&=&\frac{1}{3},\ c_{1PP}=c_{1P1P1}=\frac{3}{15},\ c_{1P0P0}=\frac{3}{5}\\
c_{P0P1}&=&-\frac{3}{15},\ c_{P1P0}=-\frac{3}{15}\\
e_{1SS}&=&\frac{1}{3},\ e_{1PP}=e_{1P0P0}=\frac{3}{15},\ e_{1P1P1}=\frac{3}{5}\\
e_{2SS}&=&\frac{1}{3},\ e_{2PP}=e_{2P1P1}=\frac{3}{15},\ e_{2P0P0}=\frac{3}{5}\\
e_{3P0P1}&=&-\frac{3}{15},\ e_{3P1P0}=-\frac{3}{15}.
\end{eqnarray}
For the $\lambda$ angular integrals, the resulting $\pm \frac{\sqrt{3}}{3}$ was already included in $H_{SP0}$ and $H_{SP1}$, respectively.
Analysing the matrix given in Eq.~(\ref{eq:bd_m}) in terms of S, P$_{\mathrm{even}}$ and P$_{\mathrm{odd}}$, one notices that the block-diagonal nature of the non-interacting terms remains preserved by the $\lambda^2$-term, only broken by mixing of S and P$_{\mathrm{odd}}$ states due to the $\lambda$-term. Note that one can show that $S$-states do not mix via the $\lambda$-term for any excited angular momentum states beyond $l=1$. However, this is not necessarily true for $\lambda^2$ contributions.
\subsubsection{Radial Integrals}

So far it was only stated that there is a matrix representation of the coupled Hamiltonian, but it was not yet specified how to treat the radial coordinates $R,\ \zeta,\ \phi$ numerically.
For this purpose, a coordinate transformation into a $h_i$-scaled perimetric coordinate system of the following form is performed in a first step, where
\begin{eqnarray}
\zeta&=&\frac{(x-y)(x+y+2z)}{4(x+y)}\\
R&=&\frac{x+y}{2}\\
\rho&=&\frac{\sqrt{xyz(x+y+z)}}{x+y}
\end{eqnarray} 
with new volume element\cite{Rene_diploma}
\begin{eqnarray}
dV=h_1 h_2 h_3\sin(\theta)(\tilde{x}+\tilde{y})(\tilde{x}+\tilde{z})(\tilde{y}+\tilde{z}) dR_{cx}dR_{cy}dR_{cz}d\tilde{x}d\tilde{y}d\tilde{z}d\phi d\theta d\psi,
\label{eq:vol2}
\end{eqnarray} 
and $\tilde{x}:=h_1 x$, $\tilde{y}:=h_2 y$, $\tilde{z}:=h_3 z$. The scaling factors $h_i$ will later be used to adjust the radial grid to the spatial extend of simulated syste.
In a next step, the orthonormal basis given in Eqs. (\ref{eq:waveansatz_first})-(\ref{eq:waveansatz_last}) is rewritten as,
 \begin{eqnarray}
\varphi_S(R,\rho,\zeta)&=&\sum_{i=1}^{N_{\mathrm{matter}}}\sum_{j=1}^{N_{\mathrm{matter}}}\sum_{k=1}^{N_{\mathrm{matter}}} N_{Sijk} F_{ijk}(\tilde{x},\tilde{y},\tilde{z})\label{eq:sm}\\
\varphi_P^{\mathrm{e}}(R,\rho,\zeta)&=&\!\!\mathcal{R}(\tilde{x},\tilde{y},\tilde{z})\!\!\sum_{i=1}^{N_{\mathrm{matter}}}\sum_{j=1}^{N_{\mathrm{matter}}}\sum_{k=1}^{N_{\mathrm{matter}}} \!\!N_{Pijk}\mathcal{R}^{-1}(h_1 x_i,h_2 y_j,h_3 z_j)^{-1} F_{ijk}(\tilde{x},\tilde{y},\tilde{z})\\
\varphi_{P0}^{\mathrm{o}}(R,\rho,\zeta)&=&\sum_{i=1}^{N_{\mathrm{matter}}}\sum_{j=1}^{N_{\mathrm{matter}}}\sum_{k=1}^{N_{\mathrm{matter}}} N_{P0ijk} F_{ijk}(\tilde{x},\tilde{y},\tilde{z})\\
 \varphi_{P1}^{\mathrm{o}}(R,\rho,\zeta)&=&\!\!\mathcal{R}(\tilde{x},\tilde{y},\tilde{z})\!\!\sum_{i=1}^{N_{\mathrm{matter}}}\sum_{j=1}^{N_{\mathrm{matter}}}\sum_{k=1}^{N_{\mathrm{matter}}} \!\!N_{P1ijk} \mathcal{R}^{-1}(h_1 x_i,h_2 y_j,h_3 z_j)F_{ijk}(\tilde{x},\tilde{y},\tilde{z}),\label{eq:P1}
\label{eq:waveansatz}
\end{eqnarray} 
where a regularization factor $\mathcal{R}(x,y,z)=\rho R=\frac{\sqrt{xyz(x+y+z)}}{2}$ was introduced  in agreement with the literature~\cite{hesse2001lagrange}. It suppresses singularities of the matter-only Hamiltonian, which may cause numerical difficulties. However, its only practical relevance is restricted to radial momentum operators, which do not appear in the coupling Hamiltonian and thus $\mathcal{R}$ eventually cancels. 
The newly introduced scaled Lagrange function $F_{ijk}(\tilde{x},\tilde{y},\tilde{z})$ is defined as,
\begin{eqnarray}
F_{ijk}(\tilde{x},\tilde{y},\tilde{z})=(N_{ijk}h_1h_2h_3)^{-1/2}\tilde{f}_i(\tilde{x}/h_1)\tilde{f}_j(\tilde{y}/h_2)\tilde{f}_k(\tilde{z}/h_2),
\end{eqnarray}
with
\begin{eqnarray}
N_{ijk}=(h_1 x_i+h_2 y_j)(h_1 x_i+h_3z_k)(h_2 y_j+h_3 z_k).
\end{eqnarray}
The Lagrange-Laguerre functions are defined as,
\begin{eqnarray}
\tilde{f}_i(u):=(-1)^i u_i^{1/2}\frac{L_{N_{\mathrm{matter}}}(u)}{u-u_i}e^{-u/2} 
\end{eqnarray}
with $L_{N}$ the Laguerre polynomial of degree N with roots $u_i$ and Lagrange property $f_i(u_j)=(\lambda_i^N)^{-1/2}\delta_{ij}$. The coefficients $\lambda_i^N$ can be chosen to fulfill the Gauss-Laguerre quadrature approximation
\begin{eqnarray}
\int_0^\infty G(u)du\approx\sum_{i=1}^N\lambda_i^N G(u_i)=\sum_{i=1}^N h \lambda_i^N G(u_i h).
\end{eqnarray}
Notice that for the formation of singlet or triplet states, the matter-only wave functions in Eqs.~(\ref{eq:sm})-(\ref{eq:P1}) can be (anti)-symmetrized by proper permutation of the perimetric coordinates (see Ref.~\citenum{hesse2001lagrange}).
Eventually, the matrix elements given in Eqs.~(\ref{eq:mel_first})-(\ref{eq:mel_last}) assume a simple form,
\begin{eqnarray}
H_{SS}&=&\delta_{ii^\prime}\delta_{jj^\prime}\delta_{kk^\prime}H_{SS}(h_1 x_i,h_2 y_j,h_3 z_k)\\
H_{PP}&=&\delta_{ii^\prime}\delta_{jj^\prime}\delta_{kk^\prime} H_{PP}(h_1 x_i,h_2 y_j,h_3 z_k)\\
H_{P0P0}&=&\delta_{ii^\prime}\delta_{jj^\prime}\delta_{kk^\prime} H_{P0P0}(h_1 x_i,h_2 y_j,h_3 z_k)\\
H_{P1P0}&=&H_{P0P1}=\delta_{ii^\prime}\delta_{jj^\prime}\delta_{kk^\prime}H_{P1P0}(h_1 x_i,h_2 y_j,h_3 z_k)\\
H_{P1P1}&=&H_{PP}=\delta_{ii^\prime}\delta_{jj^\prime}\delta_{kk^\prime} H_{P1P1}(h_1 x_i,h_2 y_j,h_3 z_k)\\
H_{SP0}&=&H_{P0S}=\delta_{ii^\prime}\delta_{jj^\prime}\delta_{kk^\prime} H_{SP0}(h_1 x_i,h_2 y_j,h_3 z_k)\\\
H_{SP1}&=&H_{P1S}=\delta_{ii^\prime}\delta_{jj^\prime}\delta_{kk^\prime}  H_{SP1}(h_1 x_i,h_2 y_j,h_3 z_k),
\end{eqnarray}
by using the orthonormality property of $F_{ijk}$ for the perimetric volume element in Gauss-approximation~\cite{hesse2001lagrange}
\begin{eqnarray}
\int_0^\infty d\tilde{x}\int_0^\infty d\tilde{y}\int_0^\infty d\tilde{z} h_1 h_2 h_3 (\tilde{x}+\tilde{y})(\tilde{x}+\tilde{z})(\tilde{y}+\tilde{z})  F_{ijk}(\tilde{x},\tilde{y},\tilde{z})F_{i^\prime j^\prime k^\prime}(\tilde{x},\tilde{y},\tilde{z})=\delta_{ii^\prime}\delta_{jj^\prime}\delta_{kk^\prime}.
\end{eqnarray}
$N_{ijk}=1$ is implied from the normalisation condition. Hence, the original eigenvalue problem given in Eq.~(\ref{eq:PFHeig}) is now discretized and numerically accessible by solving
\begin{eqnarray}
\mathcal{H}^\prime \bold{c}^\prime=E \bold{c}^\prime,
\end{eqnarray}
for $E$ and $\bold{c}$.

\subsection{Simulation Details}

For all He and HD+ simulations, the following basis set size was chosen: $N_{pt}=6$, $N_l=1$, $N_m=0$, $N_{\mathrm{matter}}=12$. For H$_2^+$ the matter grid was slightly increased and combined with a reduce photon number: $N_{pt}=5$, $N_l=1$, $N_m=0$, $N_{\mathrm{matter}}=16$. Therefore, for each input parameter combination a Hamiltonian matrix of size $41472^2$ for distinguishable particles (HD+), $21600^2$ for He and $42240^2$ for H$_2^+$   had to be  diagonalized. The eigenvalue problem was implemented in the in-house LIBQED python code and the high-performance ELPA library~\cite{marek2014elpa} was used for the exact numerical diagonalization.

The particle masses were set according to literature~\cite{hesse2001lagrange,alexander1988high}, i.e.
He: $m_1=1$, $m_2=1$, $m_3=7294.2618241$, HD+: $m_1=1836.142701$, $m_2=3670.581$, $m_3=1$ and H$_2^+$: $m_1=1836.142701$, $m_2=1836.142701$, $m_3=1$.
Corresponding scaling values for the radial Lagrange-Laguerre grid were set to the following values, which were motivated by matter-only considerations in the literature~\cite{Rene_diploma}:
He: $h_1=0.8$, $h_2=0.8$, $h_3=0.4$, HD+: $h_1=0.16$, $h_2=0.16$, $h_3=1.4$ and H$_2^+$: $h_1=0.33$, $h_2=0.33$, $h_3=2.0$.
The different scaling values account for the difference in the spatial localisation of the constituents and thus allow to reach high numerical accuracy with a relatively coarse radial grid.

\subsection{Convergence and Numerical Tests}

Multiple explicit convergence/ sanity checks were perform to ensure that finite basis set errors or implementation mistakes do not spoil the results. Matter-only energy eigenvalues were compared with reference calculations from literature given in Tab.~\ref{tab:en}. For He, the $\lambda^2$-scaling of the ground-state could be compared with QEDFT calculations with photon OEP accuracy in the Born-Oppenheimer limit, which indicates an agreement on the same accuracy level as one expects from the previous matter-only considerations (see Fig.~\ref{tab:en}). All QEDFT simulations were performed with the OCTOPUS code~\cite{tancogne2020octopus}. Note that highly accurate simulations of ground-state nuclear contributions in QEDFT would be very challenging to obtain~\cite{flick2018cavity}.

Last but not least, the Thomas-Reiche-Kuhn sum rule was well preserved for He, HD+ and H$_2^+$, which implies that there is no fundamental implementation error present. Moreover, all observables were consistent with theory and particularly in agreement with the JC model. This offers an additional sanity check of the numerical results (see main section of the manuscript).

\begin{table}
\begin{tabular}{ |p{2.5cm}|p{3.4cm}|p{3.4cm}|   }
 \hline
 \multicolumn{3}{|c|}{Lowest matter-only energy eigenvalues [H]} \\
 \hline
 \hline
He & Reference \cite{hesse1999lagrange,hesse2001lagrange} & $N_{\mathrm{matter}}=12$ \\
 \hline
 1S$^{\mathrm{(even)}}$   & -2.9033045555597   & -2.90330437154 \\
 2S$^{\mathrm{(even)}}$ &   -2.145678586051 &   -2.14567817793 \\
 2P$^{\mathrm{(odd)}}$  & -2.1235456525895 &          -2.12320490110 \\
 
\hline
\hline
 HD+ & Reference \cite{korobov2006leading} & $N_{\mathrm{matter}}=12$\\
\hline
1S$^{\mathrm{(even)}}$& -0.59789796860903 &  -0.59757212 \\
  2P$^{\mathrm{(odd)}}$ &  -0.59769812819221 & -0.59737196 \\
 2S$^{\mathrm{(even)}}$& -0.58918182955696 & -0.58702708 \\

\hline
\hline
 H$_2^+$ & Reference \cite{Rene_diploma} & $N_{\mathrm{matter}}=16$\\
\hline
 1S$^{\mathrm{(even)}}$ & -0.597139063121 &  -0.596973\\
  2S$^{\mathrm{(even)}}$& -0.58715567914 & -0.585948 \\
 2P$^{\mathrm{(odd)}*}$ & -0.4990065652928$^*$ & -0.498039$^*$\\
 \hline
\end{tabular}
 \caption{Despite having a substantially smaller radial matter grid available for our coupled simulations compared with matter-only reference data, our grid allows to reach millihartree accuracies for the absolute energy eigenvalues and orders of magnitude smaller values for corresponding energy-differences. (*) Notice that the  2P$^{\mathrm{(odd)}}$ state for H$_2^+$ corresponds to the dissociation limit~\cite{Rene_diploma}. In other words, there are no dipole allowed bound state transitions for H$_2^+$ and a continuum of allowed transitions arises beyond this energy value.}
 \label{tab:en}
\end{table}

\begin{figure}[H]
    \centering
    \includegraphics[width=0.8\linewidth]{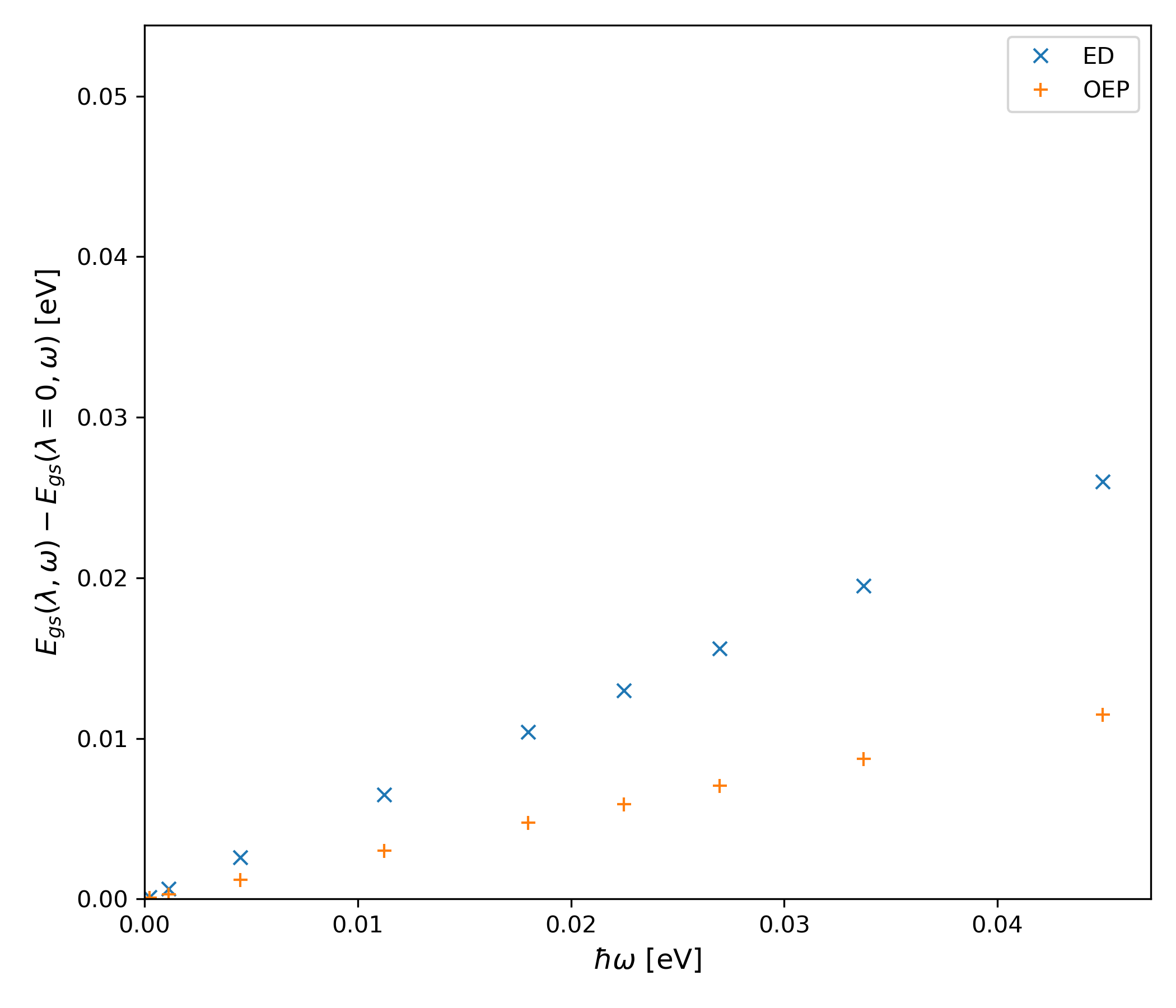}
    \caption{Cutting the basis set expansion for angular momentum quantum numbers $l>1$, may introduce significant numerical errors for stronger couplings. In order to check the validity by allowing e.g. $l=2$, the code complexity of the implementation would be more than doubled. Therefore, we decided to use an alternative route. The comparison of the the He ground-state energy shift with results from QEDFT simulations with photon OEP \cite{flick2018ab} indicates that inaccuracies from $l<2$, are on the order of milihartree or below, which is in line with the accuracy reached for the absolute matter-only energies given in Tab. \ref{tab:en}. As it is the case for matter-only values, one expects considerable smaller relative errors in terms of energy differences. }
    \label{fig:gs_oep_ed}
\end{figure}

\section{Results: Additional Observables}

\subsection{He}
\begin{figure}[H]
    \begin{subfigure}{.49\textwidth}
    \centering
    \includegraphics[width=1\linewidth]{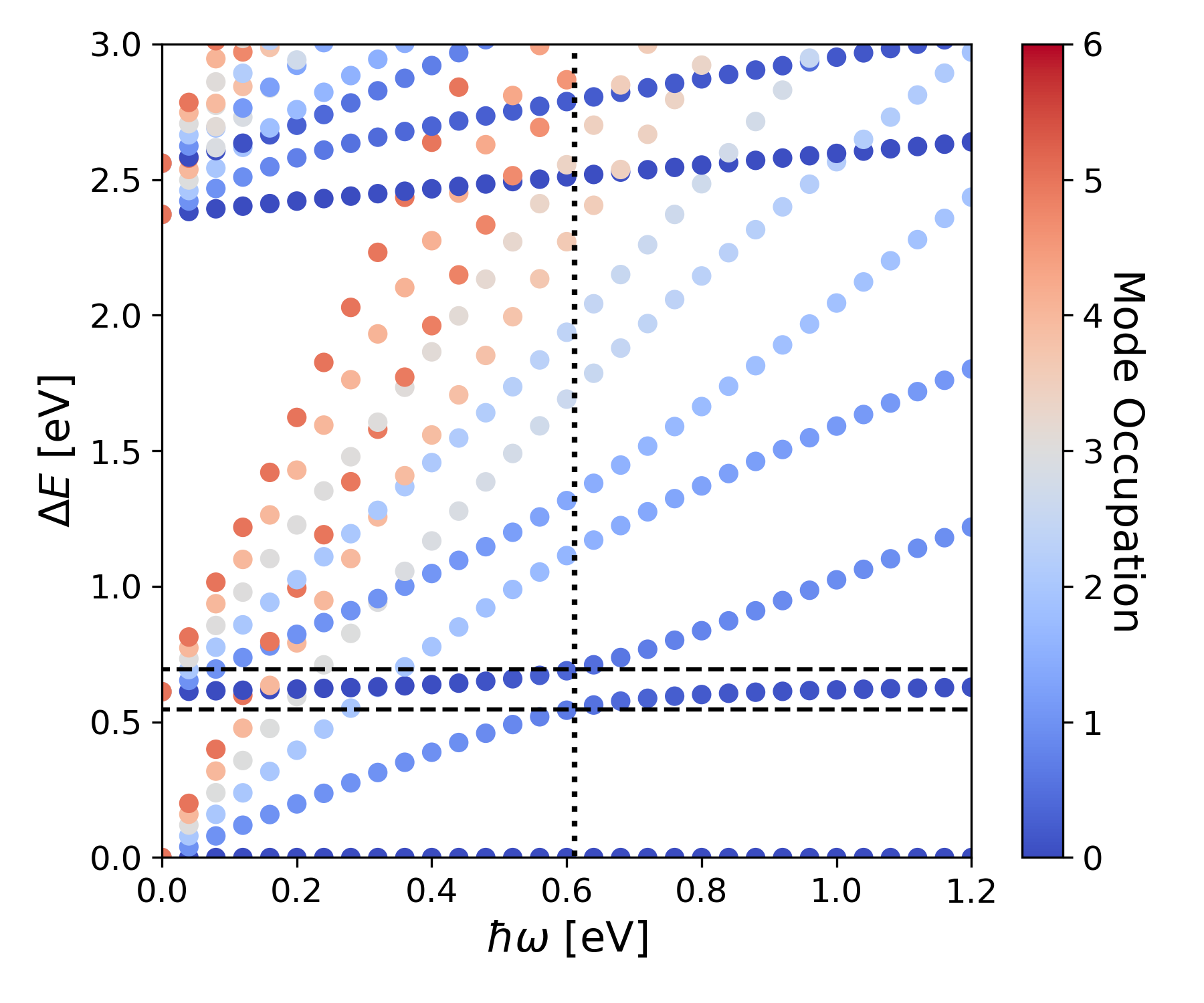}
    \caption{}
    \label{fig:he_n}
\end{subfigure}
\begin{subfigure}{.49\textwidth}
    \centering
    \includegraphics[width=1\linewidth]{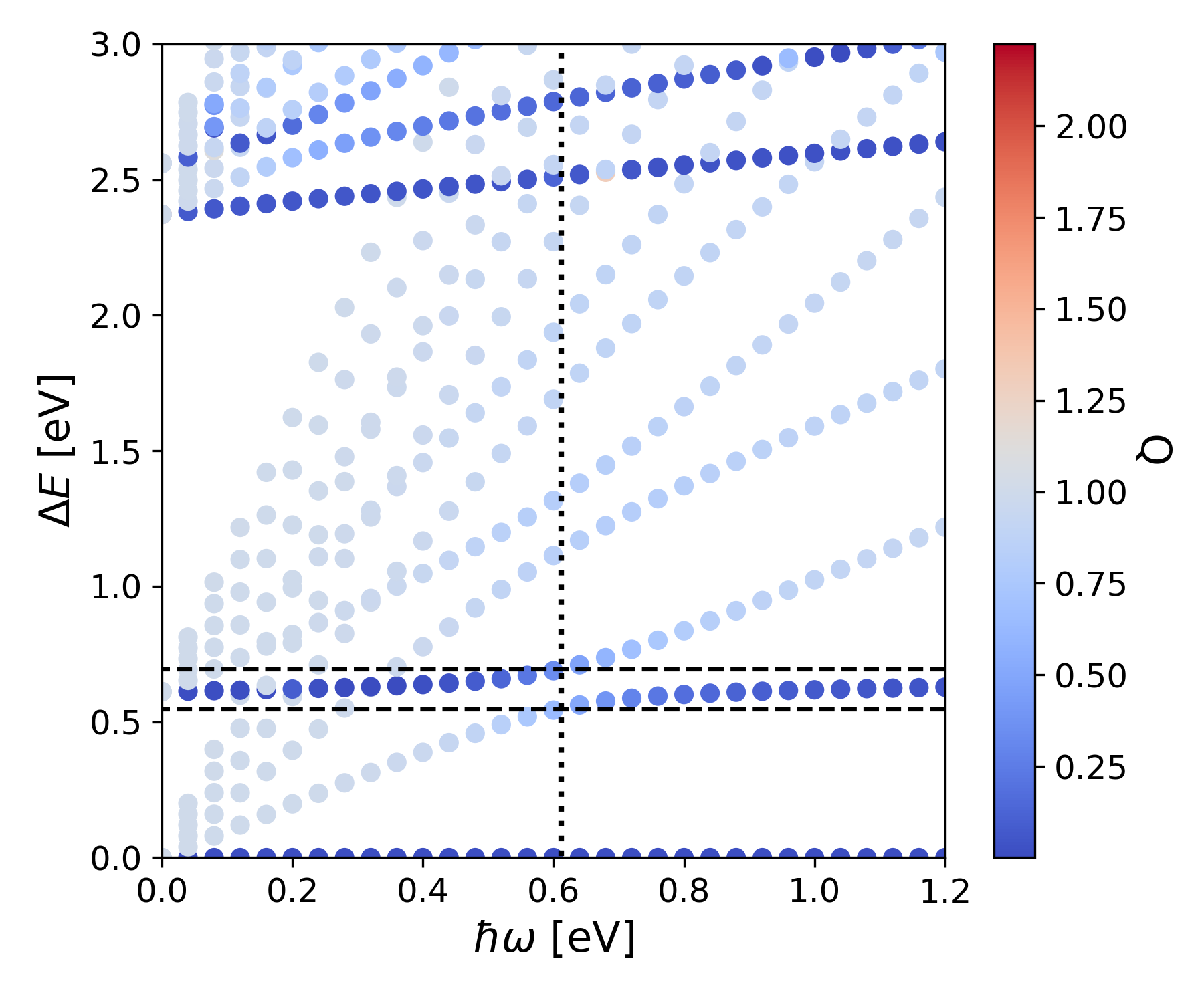}
    \caption{}
    \label{}
\end{subfigure}
\begin{subfigure}{.49\textwidth}
    \centering
    \includegraphics[width=1\linewidth]{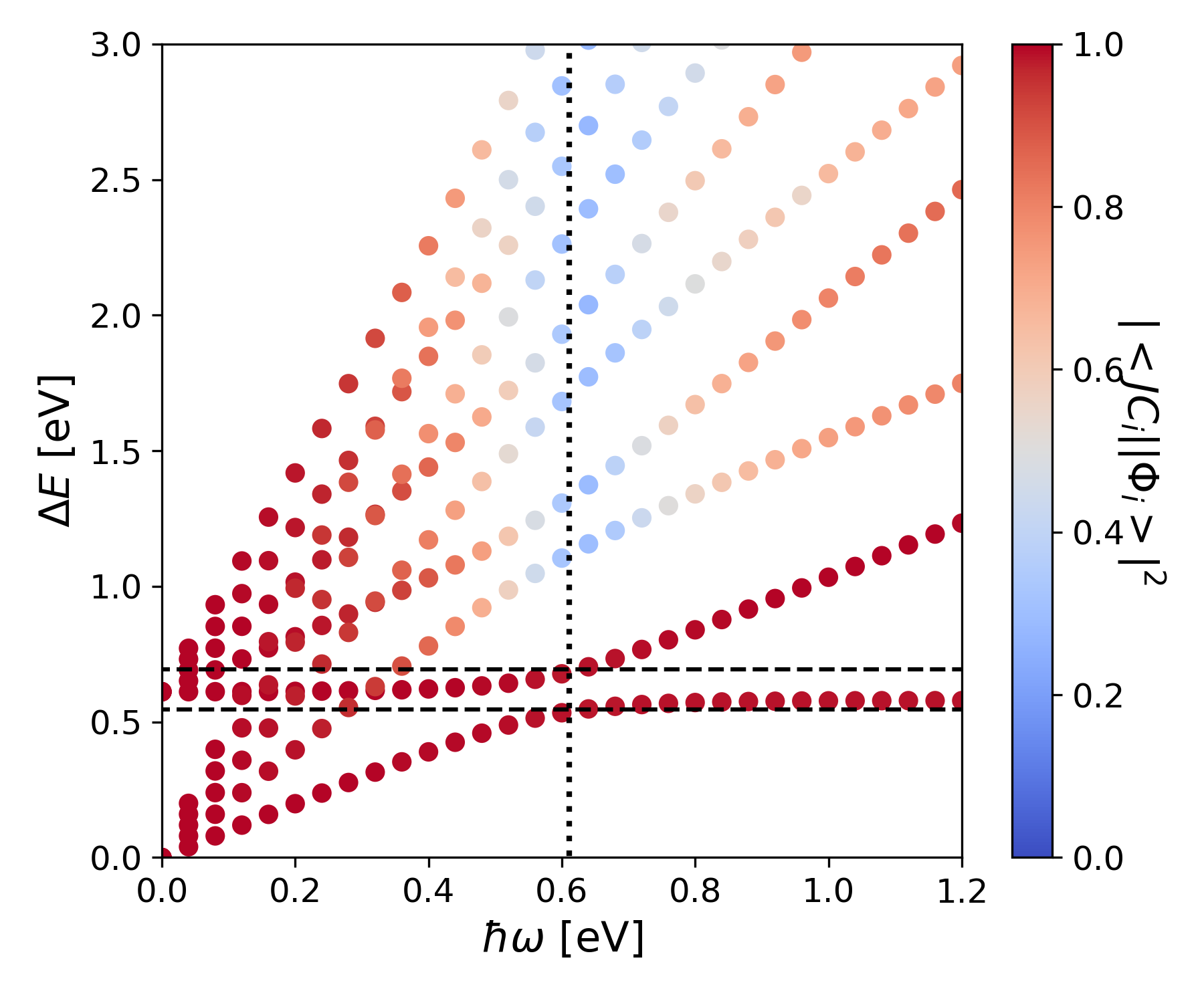}
    \caption{}
    \label{fig:he_overlap}
\end{subfigure}
\caption{Parahelium polaritonic dispersion curves in a cavity. The vertical line indicates the $2$S-$2$P resonance where $\lambda=0.057$ was set. Horizontal lines indicate the splitting of the lowest two polaritons. Notice that $\frac{\sqrt{\omega}}{\lambda}$ was kept fix, i.e. the coupling strength $g\propto\omega\hbar$. In (a) and (b) the color bars indicates photonic observables $\langle \hat{n}\rangle$ and Mandel $Q$-parameter respectively. Whereas, in (c) the wave-function overlap between our exact calculation and the corresponding JC model is shown. }
\end{figure}

Figs.~\ref{fig:he_n} - \ref{fig:he_overlap} show the parahelium dispersion curves with respect to the mode occupation $\langle \hat{n}\rangle=\langle\hat{a}^\dagger \hat{a}\rangle$, Mandel $Q$-parameter $Q:=\tfrac{\langle\hat{a}^\dagger\hat{a}^\dagger\hat{a}\hat{a}\rangle-\langle\hat{a}^\dagger\hat{a}\rangle^2}{\langle\hat{a}^\dagger\hat{a}\rangle}$ and wave-function overlap between the exact solution and the JC model. Notice, that the $\Delta E$-values in the last figure are obtained from the JC model (i.e. based on matter-only considerations) and not from the exact diagonalisation of the coupled system. The mode occupation (a) clearly highlights the one- to five-photon lines (replica) that appear in our simulations (we have chosen $N_{pt}=6$ in these simulations). With each further Fock-basis state in our simulation we would get a further photon line. The Mandel $Q$-parameter (b) indicates the nature of the photon subsystem. If $Q<0$ we would have a Fock-like state while for $Q\geq 0$ we have a classical (or even chaotic) photon state. In our case we have mainly classical photon states. Finally, in (c) we see that for the standard upper and lower polaritons the JC model is highly accurate even on the wave-function level, while the higher photon-replicas are not well captured at resonance.

\subsection{HD+}
\begin{figure}[H]
\begin{subfigure}{.49\textwidth}
    \centering
    \includegraphics[width=1\linewidth]{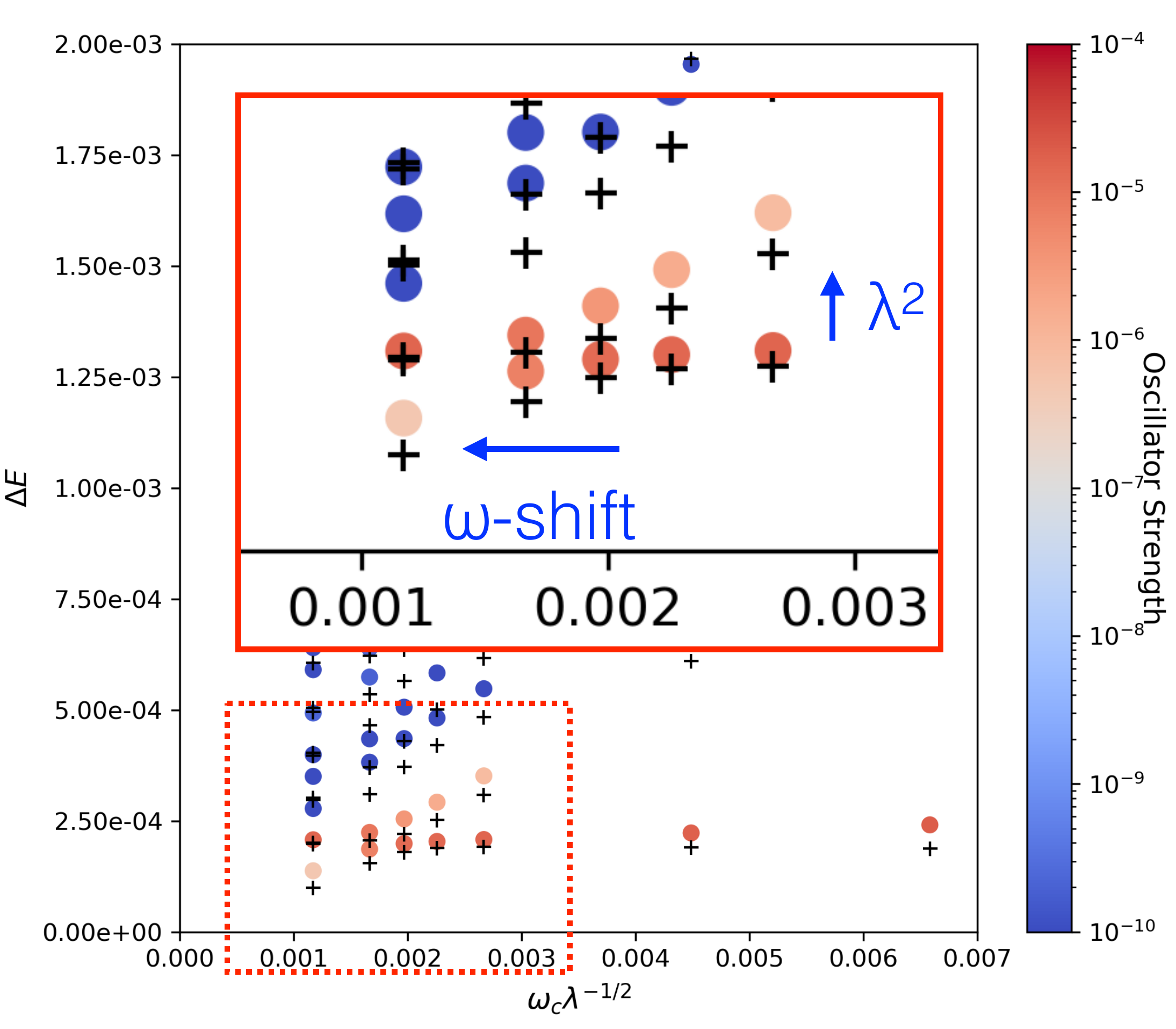}
   
\end{subfigure}
\caption{Visualization of the dressed polaritonic dispersion relation of HD+ in a cavity with a frequency centered around the fundamental ro-vibrational transition in atomic units. The shifts are caused by dipole self-interaction contributions ($\Delta E$-shift) and COM influence for non-zero net-charge $\hbar\omega$-shift.   }
 \label{fig:hdp_zoom}
\end{figure}

\begin{figure}[H]
\begin{subfigure}{.49\textwidth}
\centering
    \includegraphics[width=1\linewidth]{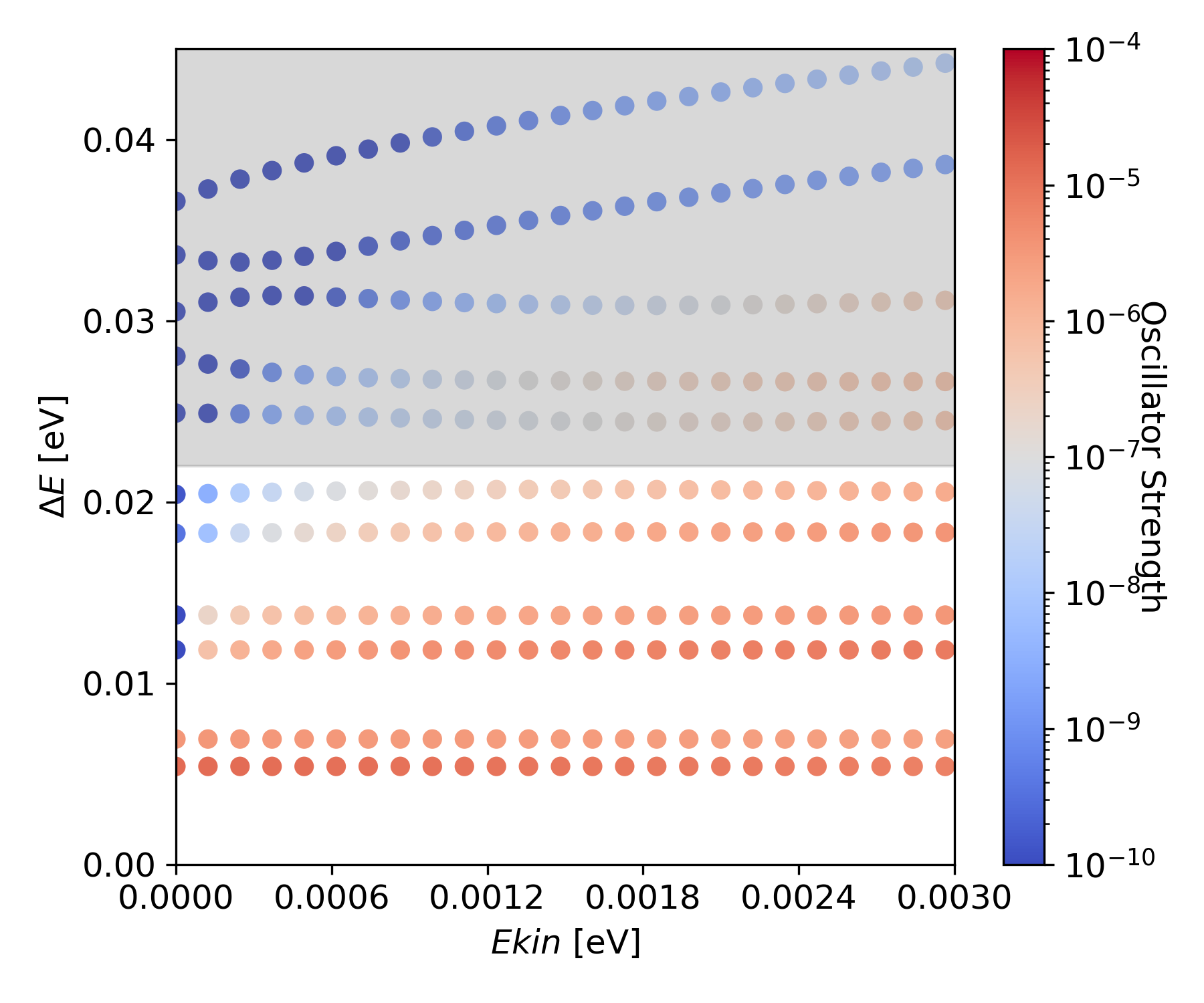}
    \caption{}
    \label{fig:E_lambda_HDp_k1}
\end{subfigure}
\begin{subfigure}{.49\textwidth}
\centering
    \includegraphics[width=1\linewidth]{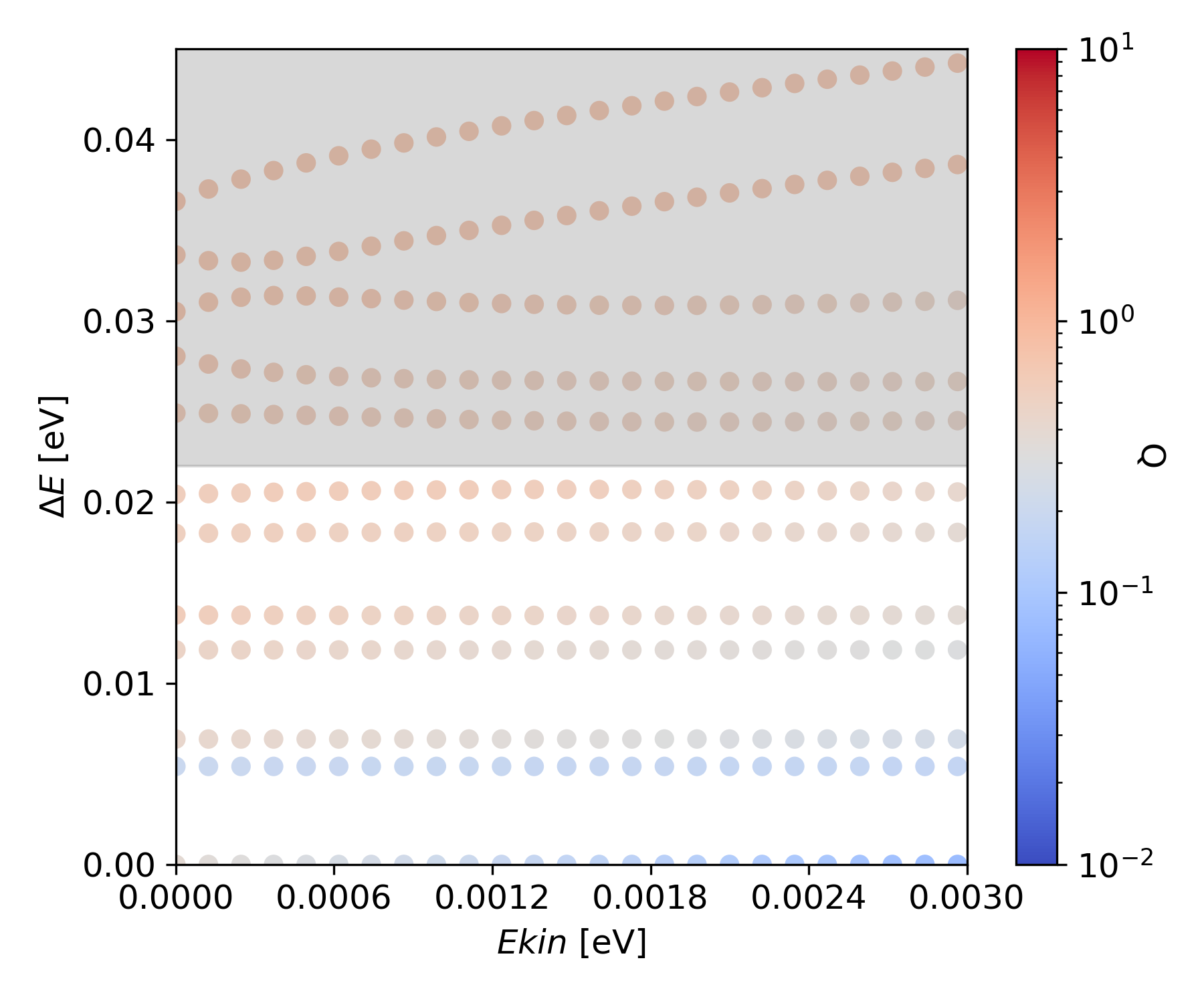}
    \caption{}
    \label{fig:E_lambda_HDp_k1_Q}
\end{subfigure}
\begin{subfigure}{.49\textwidth}
    \centering
    \includegraphics[width=1\linewidth]{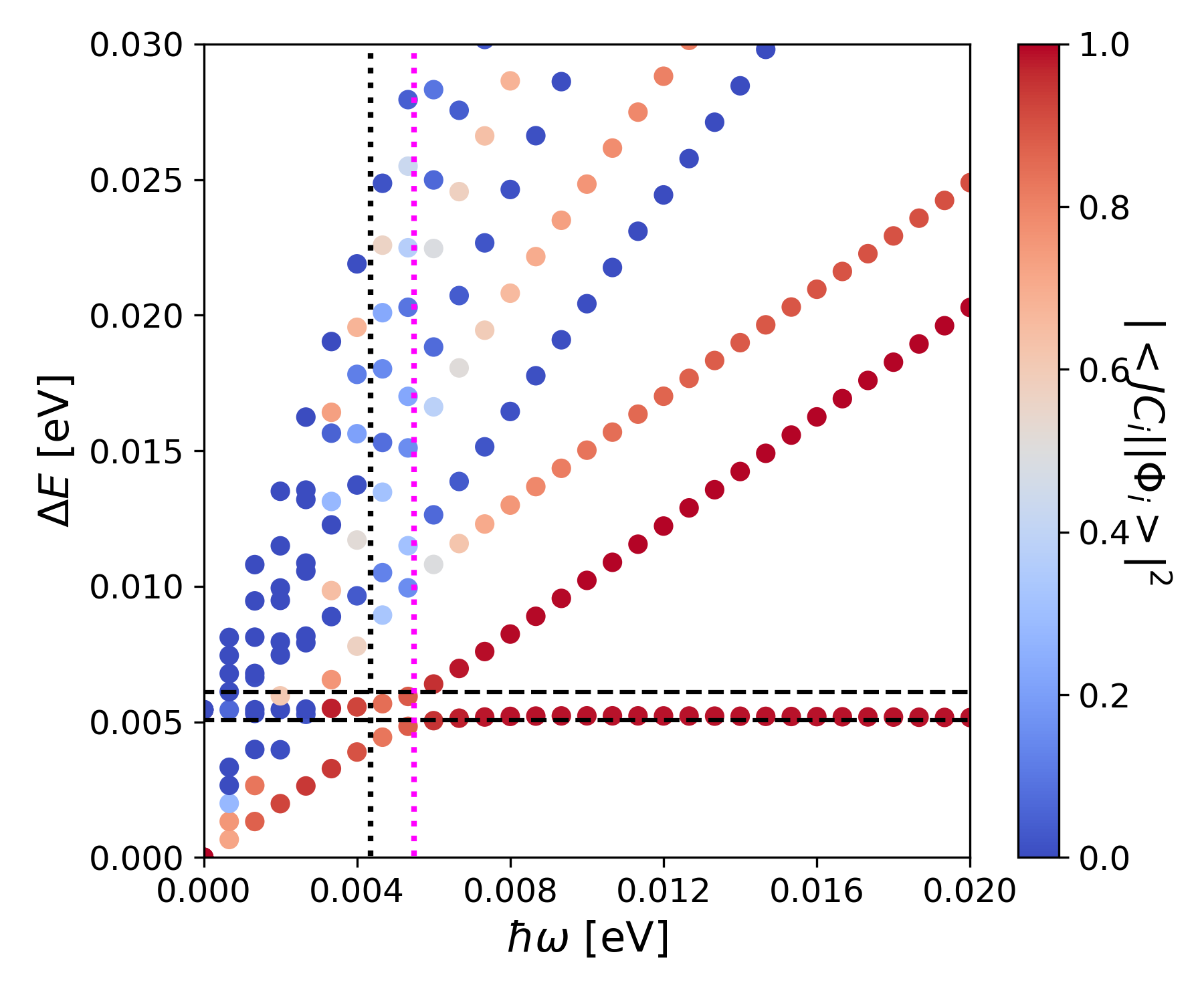}
    \caption{}
    \label{fig:hdp_o_k0}
\end{subfigure}
\begin{subfigure}{.49\textwidth}
    \centering
    \includegraphics[width=1\linewidth]{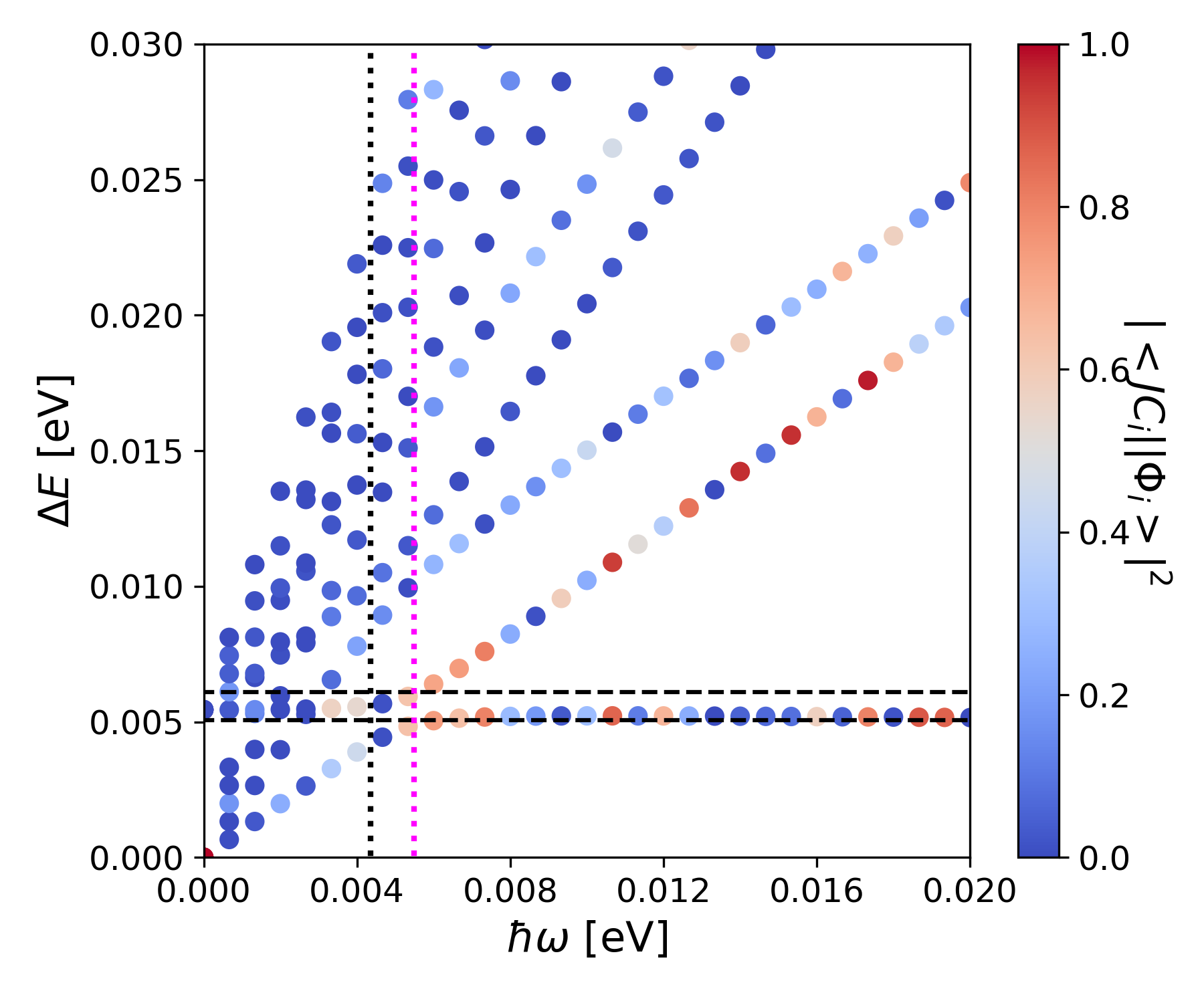}
    \caption{}
    \label{fig:hdp_o_k1}
\end{subfigure}
\caption{ In (a) and (b) we consider the HD+ resonant case for $\Delta E$ at $\lambda=0.01$ with respect to the kinetic energy $E_{kin}$ of the COM. While the spectrum is not changed (up to numerical inaccuracies for higher-lying states) the COM motion (a) redistributes the oscillator strengths and (b) also modifies the properties of the photons. Here a reduction of the Mandel $Q$-parameter indicates the photons to be in almost a coherent state. Notice that the grey area indicates less reliable eigenvalues, which are not converged for the chosen photon number basis with $N_{pt}=6$. 
In (c) and (d) the HD+ polaritonic dispersion curves for $k_z=0$ and $k_z=1$ are shown with respect to the wave-function overlap between our exact calculations and the corresponding JC model. The black vertical line indicates the $2$S-$2$P resonance condition $\hbar\omega$ and the purple vertical line indicates the corresponding frequency predicted by the JC model that is missing the frequency dressing. Notice that the energy eigenvalues shown in (c) and (d) are determined by the JC model and not by the exact diagonalisation of the coupled problem. Horizontal lines indicate the splitting of the lowest two polaritons.
}
\end{figure}

In Fig.~\ref{fig:hdp_zoom} a zoom of the dispersion relations given in the main section is shown to visualize dressing effects, caused by dipole self-interaction, and the shift of the resonance frequency due to the non-zero net-charge. 
In Figs.~\ref{fig:E_lambda_HDp_k1} and \ref{fig:E_lambda_HDp_k1_Q} the impact of different finite COM velocities on the oscillator strength and Mandel $Q$-parameter is shown at resonance condition.
The break-down of the JC model at finite COM velocities for charged systems is visualised in Figs.~\ref{fig:hdp_o_k0} and \ref{fig:hdp_o_k1}. They indicate that the relatively high agreement between exact and JC wave function for bright states at $k_z=0$ breaks down at finite velocities. For such systems one expects for any observable, which is calculated from JC states, to be error-prone at finite COM velocities.


\subsection{H$_2^+$}

\begin{figure}[H]
\begin{subfigure}{.49\textwidth}
\centering
    \includegraphics[width=1\linewidth]{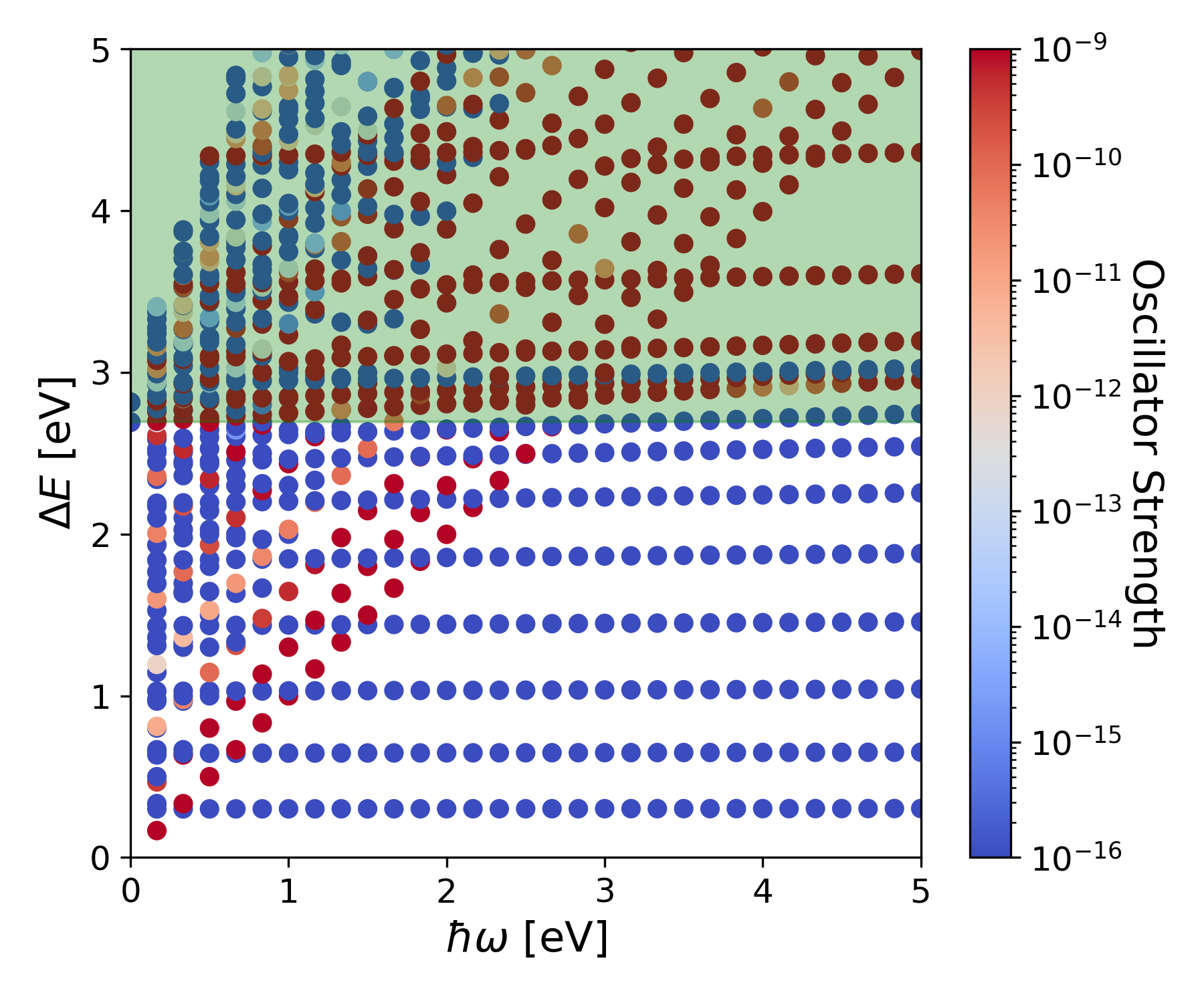}
    \caption{}
    \label{fig:disp_H2p_k0}
\end{subfigure}
\begin{subfigure}{.49\textwidth}
\centering
    \includegraphics[width=1\linewidth]{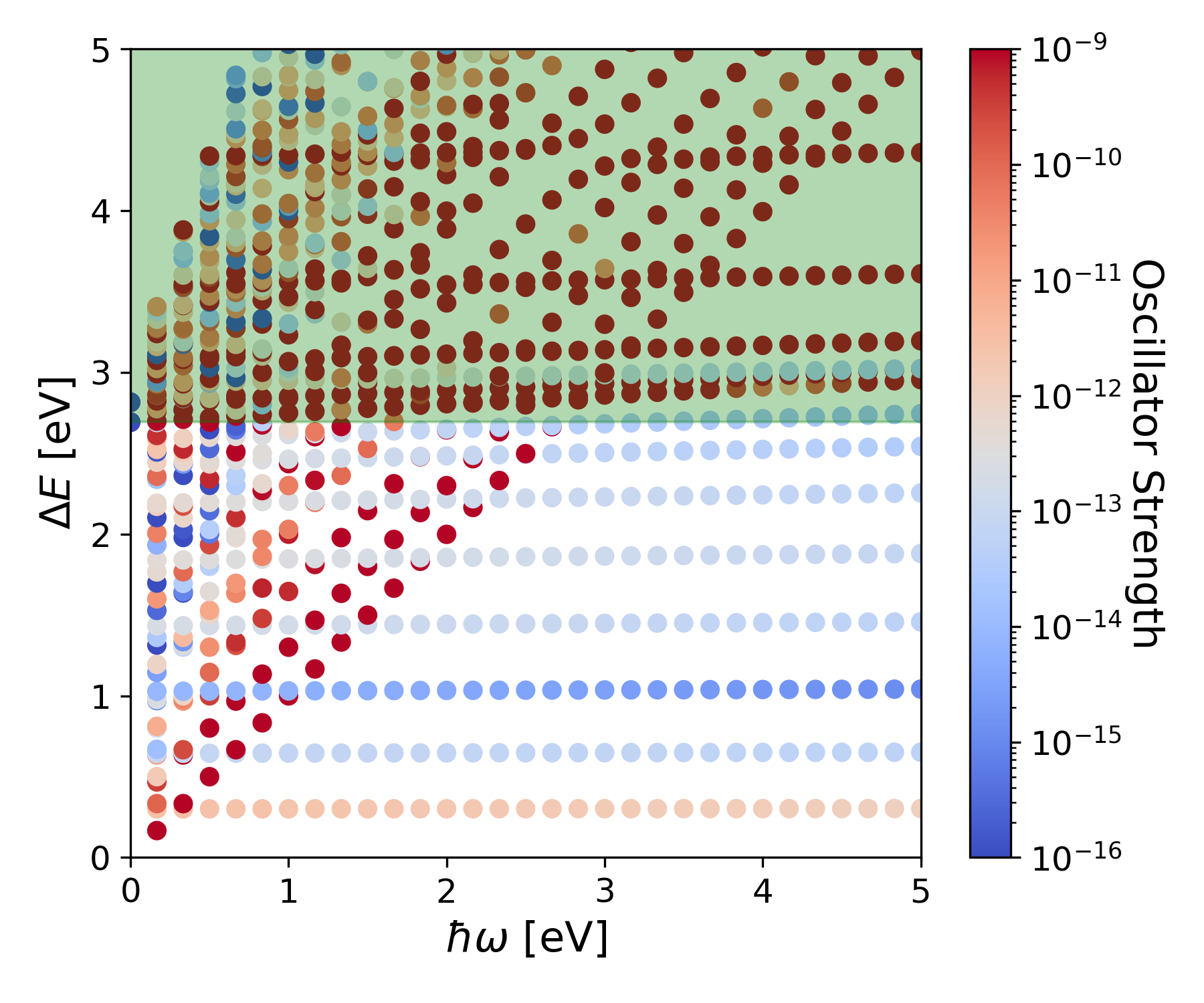}
    \caption{}
    \label{fig:disp_H2p_k1}
\end{subfigure}
\caption{Dispersion relations for the H$_2^+$ molecule with singlet nuclear spin configuration in a cavity. The frequencies are centered around the dissociation energy of the bare matter system. It is assumed that $\frac{\sqrt{\hbar \omega}}{\lambda}=const$, i.e. the coupling strength $\propto \omega$, and $\lambda=0.057$ at resonance with the dissociation energy. The oscillator strength color bar is chosen to visualize changes arising from finite COM velocities in a relatively weak regime. The COM motion was set to $E_{kin}=0$ for (a), whereas for (b) a non-vanishing $E_{kin}=0.37$ [eV] was chosen, which is still in the non-relativistic limit, i.e. $k\approx 0.072 c$.
The vertical dotted line indicates the dissociation resonance condition. }
\end{figure}

\begin{figure}[H]
    \centering
    \includegraphics[width=0.7\linewidth]{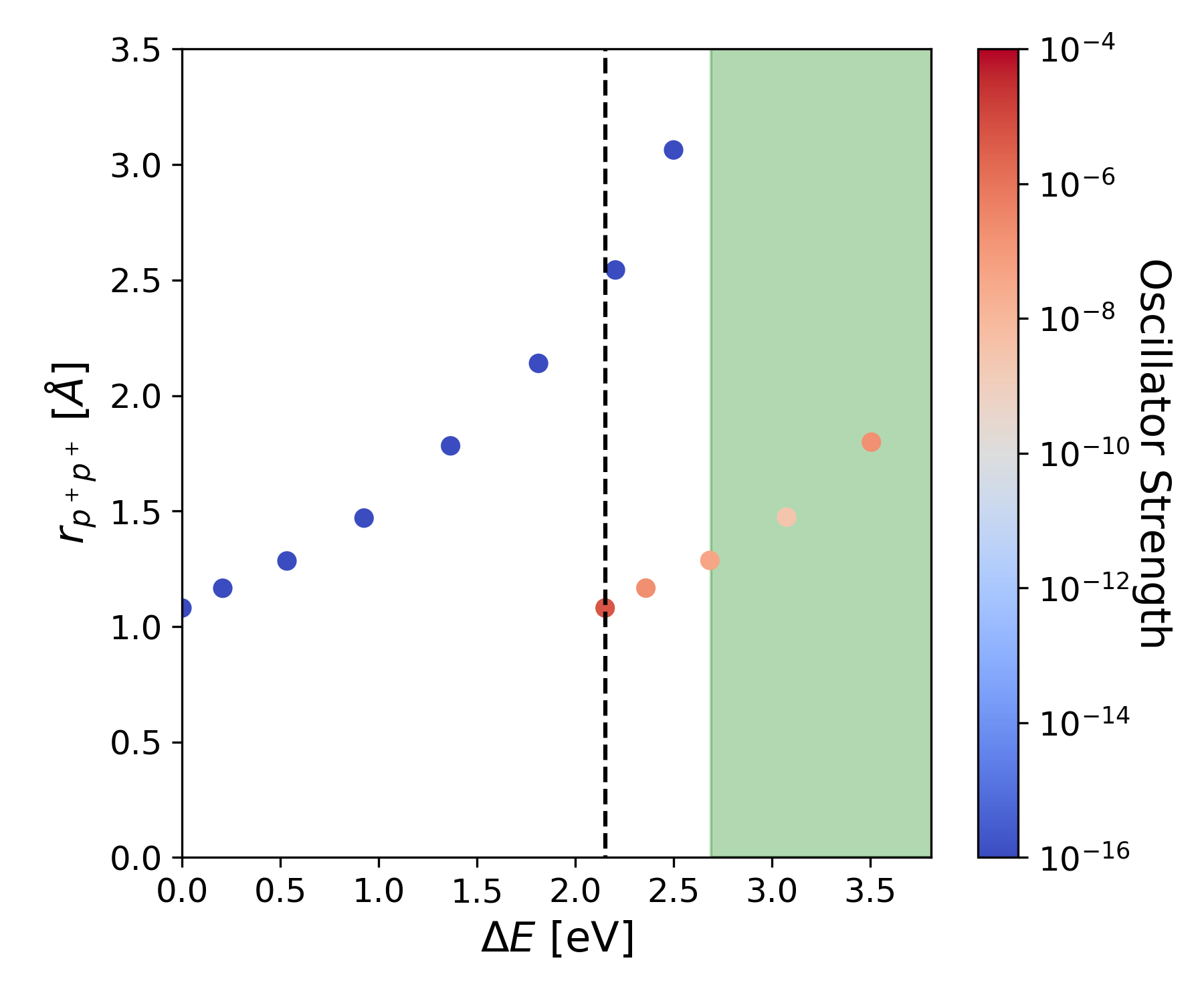}
    \caption{Quantized (i.e. bound) proton-proton distances for H$_2^+$ with respect to ground-state energy differences $\Delta E$ and corresponding oscillator strength in the  \textit{infinite proton mass limit}. The blue dots correspond to dressed bare matter states whereas red dots indicate the emerging bright photon replicas, which are absent without a cavity. The cavity frequency is $\omega=2.15$ eV (dashed vertical line) with $\lambda=0.051$ and zero COM motion. The green area indicates energy ranges beyond the p$^+$-dissociation limit according to matter-only simulations.}
    \label{fig:dist_H2p_BO}
\end{figure}

Similarly to HD+, we performed simulations for H$_2^+$ at different COM velocities (see Figs.~\ref{fig:disp_H2p_k0} - \ref{fig:disp_H2p_k1}). In contrast to HD+, the frquencies are scanned around the $1$S-$2$P transition, which corresponds to the dissociation limit H$_2^+\rightarrow$ H $+$ p. This adds additional complexity to the interpretation of the computed dispersion relations. One needs to consider that a continuum of dipole-allowed transitions emerges beyond the dissociation limit. However, this cannot be represented on our finite radial grid, which is scaled to reproduce bound-state properties optimally. For this reason, one observes a discrete spectrum of dipole allowed (red) energy levels beyond the dissociation limit. However, there are two ways to identify truly discrete (i.e. bound) peaks in our discretized continuum. First, one can calculate the proton-proton distances for each excited state to identify potentially bound states (see main section). Second, bound excitations are invariant with respect to changes of the radial basis set and the corresponding scaling parameters, whereas the discrete continuum reacts very sensitively. Based on these considerations, we could distinguish the bound states, which are shown in the main section, from continuum states composed of a H-atom and a free proton coupled to the cavity. With the latter method one can also identify the dissociation energy limits, i.e. when increasing the scaling factors of the radial grids one observes an accumulation of eigenvalues at the dissociation energies~\cite{Rene_diploma}, while the newly discovered bound polaritonic states remain invariant.
Similarly to the main section, in Fig.~\ref{fig:dist_H2p_BO} the proton-proton distance vs. energy plot is shown for the previously identified bound states. However, for this figure we assumed an infinite mass for the nuclei. Qualitatively, the system behaves very similar compared with the results for finite proton masses. Overall, the proton-proton distances of the bound states are slightly reduced and some of the excitation energy differences are moderately shifted compared with the finite mass reference. Nevertheless, the appearance of bound states below and above the dissociation limit remains preserved in the infinite-mass limit.

Aside from that, one can also investigate the bound $1$S-$i$S transitions below the dissociation limit with respect to different COM velocities. Our simulations for H$_2^+$ confirm that a finite COM motion of a charged molecule indeed leads to an increased dipole-transition probability, as it was discussed for HD+ in the main section.

\bibliography{si}